\documentclass[prb,twocolumn,aps,10pt]{revtex4-1}

\usepackage{amsmath,amssymb}

\usepackage{graphicx}

\usepackage{bbm}

\usepackage[T1]{fontenc}

\newcommand{\ii}{\mathrm{i}}
\renewcommand{\vec}[1]{\mathbf{#1}}

\begin{document}

\title{Topological invariants for Floquet-Bloch systems \\ with chiral, time-reversal, or particle-hole symmetry}

\author{Bastian H{\"o}ckendorf}
\author{Andreas Alvermann}
\email{alvermann@physik.uni-greifswald.de}
\thanks{Author to whom any correspondence should be addressed.}
\author{Holger Fehske}
\affiliation{Institut f{\"u}r Physik, Ernst-Moritz-Arndt-Universit{\"a}t Greifswald, 17487 Greifswald, Germany}

\begin{abstract}
We introduce $\mathbb Z_2$-valued bulk invariants for symmetry-protected topological phases in $2+1$ dimensional driven quantum systems.
These invariants adapt the $W_3$-invariant, expressed as a sum over degeneracy points of the propagator,
to the respective symmetry class of the Floquet-Bloch Hamiltonian.
The bulk-boundary correspondence that holds for each invariant relates a non-zero value of the bulk invariant to
the existence of symmetry-protected topological boundary states.
To demonstrate this correspondence we apply our invariants to a chiral Harper, time-reversal Kane-Mele, and particle-hole symmetric graphene model with periodic driving,
where they successfully predict the appearance of boundary states that exist despite the trivial topological character of the Floquet bands.
Especially for particle-hole symmetry, combination of the $W_3$ and the $\mathbb Z_2$-invariants
allows us to distinguish between weak and strong topological phases.
\end{abstract}

\maketitle

\section{Introduction}

Topological states of matter~\cite{PhysRevLett.45.494,PhysRevLett.49.405,PhysRevLett.95.146802,RevModPhys.82.3045,Konig766,PhysRevLett.98.106803} have become the subject of intensive research activities over the past decade. More recently, unconventional topological phases in periodically driven systems~\cite{PhysRevB.82.235114, Floquettopological, PhysRevB.84.235108, Flaschner, PhysRevX.3.031005, PhysRevLett.116.176401} have moved into focus.
Driving allows for non-trivial topological phases even if each individual Floquet band is topologically trivial.
These phases cannot be characterized by static invariants, such as the Chern numbers of the Floquet bands, but only through invariants that depend on the entire dynamical evolution of the system~\cite{PhysRevX.3.031005}.
Irradiated solid state systems \cite{Wang453, PhysRevB.91.241404, Wang_SR} and photonic crystals \cite{nphoton.2014.248, nature12066, anomalous, anomalous_2}, where the third spatial dimension represents the time axis, are promising candidates for the realization of these new topological phases.

The relevant topological invariant of driven 2+1 dimensional systems is the $W_3$-invariant of unitary maps~\cite{PhysRevX.3.031005},
which is evaluated for the Floquet-Bloch propagator $U(\vec k,t)$
that solves the Schr\"odinger equation $\ii \partial_t U(\vec k,t) = H(\vec k,t) U(\vec k,t)$
with a periodic Hamiltonian $H(\vec k, t+T) = H(\vec k, t)$. 
The bulk-boundary correspondence for the $W_3$-invariant guarantees that the value of $W_3(\epsilon)$ equals the number of chiral boundary states in the gap around the quasienergy $\epsilon$.

The situation changes again for driven systems with additional symmetries.
Symmetry-protected boundary states appear in pairs of opposite chirality, such that the $W_3$-invariant can no longer characterize the non-trivial topological phases~\cite{PhysRevLett.112.026805, PhysRevB.90.195419, PhysRevB.90.205108, Zhou2014, PhysRevLett.114.106806, 1367-2630-17-12-125014, PhysRevB.93.075405,PhysRevB.96.155118,PhysRevB.96.195303}.
Two questions arise immediately: Can the phases be characterized by new invariants? 
Can these invariants be computed for complicated Hamiltonians and Floquet-Bloch propagators?

In this paper we try to answer both questions affirmatively
by deriving and evaluating $\mathbb Z_2$-valued bulk invariants for Floquet-Bloch systems with chiral, time-reversal, or particle-hole symmetry.
In each case, the symmetry is given by a relation of the form $H(\vec k, t) = \pm S H(\hat{\vec k}, \pm t) S^{-1}$ for the time-dependent Bloch Hamiltonian $H(\vec k,t)$,
with a (anti)-unitary operator $S$ and an involution $\vec k \mapsto \hat{\vec k}$ on the Brillouin zone $\mathcal B$.  
The symmetry relation implies a zero $W_3$-invariant in certain gaps,
because the degeneracy points of $U(\vec k, t)$ that contribute to $W_3(\epsilon)$ occur in symmetric pairs and cancel.
Conceptually, the new symmetry-adapted invariants count only one partner of each pair of degeneracy points.
Since the result depends on which partner is counted, the new invariants are $\mathbb Z_2$-valued.
Symmetry-protected topological boundary states appear in gaps where the symmetry relation enforces  $W_3(\epsilon)=0$, but the $\mathbb Z_2$-invariants are non-zero.

Topological invariants for Floquet-Bloch systems with and without additional symmetries have been introduced before~\cite{PhysRevB.82.235114, PhysRevX.3.031005, 1367-2630-17-12-125014, PhysRevLett.114.106806, PhysRevB.93.075405, PhysRevB.93.115429, PhysRevB.96.195303}, and our constructions resemble some of them~\cite{1367-2630-17-12-125014, PhysRevLett.114.106806, PhysRevB.93.075405} in various aspects.
However, most constructions in the literature differ for each symmetry. 
One of our goals is to show that the construction of invariants in terms of degeneracy points of $U(\vec k, t)$ applies to each symmetry equally, with only the obvious minimal modifications. In this way the constructions  described here constitute a unified approach to topological invariants in Floquet-Bloch systems with symmetries.

Our presentation begins in Sec.~\ref{sec:W3} with the discussion of an expression for the $W_3$-invariant that is particularly well suited for the following constructions, before the different invariants for chiral, time-reversal, and particle-hole symmetry are introduced in Sec.~\ref{sec:Z2}.
Sec.~\ref{sec:Conc} summarizes our conclusions, and the appendices (App.~\ref{app:W3}--App.~\ref{app:PH}) give details on the derivations in the main text.

\section{$W_3$-invariant and degeneracy points}
\label{sec:W3}

The starting point for the construction of the $\mathbb Z_2$-invariants is the expression
 \begin{equation}\label{W3}
W_3(\epsilon)=\sum_{\nu=1}^n \sum_{i=1}^{\mathrm{dp}} N^{\nu}(\epsilon, \vec d_i) \, C^{\nu}(\vec d_i)
\end{equation}
of the $W_3$-invariant as a sum over all degeneracy points $i = 1,\dots, \mathrm{dp}$ of the Floquet-Bloch propagator $U(\vec k,t)$
that occur during time evolution $0 \le t \le T$.

Eq.~\eqref{W3} is a modified version of an expression for $W_3(\epsilon)$ given in Ref.~\onlinecite{HAF17}.
As explained in App.~\ref{app:W3}, which contains a detailed derivation,
it generalizes a similar expression introduced in Ref.~\onlinecite{1367-2630-17-12-125014}.
A unique feature of Eq.~\eqref{W3} is the invariance of all quantities under general shifts $\epsilon(\cdot) \mapsto \epsilon(\cdot) + 2 \pi m$ of the Floquet quasienergies, whereby the ambiguity of mapping eigenvalues $e^{-\ii \epsilon(\cdot)}$ of $U(\cdot)$ to quasienergies $\epsilon(\cdot)$ is resolved from the outset.
For this reason, Eq.~\eqref{W3} is 
particularly convenient for numerical evaluation, e.g., with the algorithm from Ref.~\onlinecite{HAF17}.
Note that for the sake of clarity of the main presentation we assume in Eq.~\eqref{W3} that the bands are topologically trivial for $t \to 0$.
The general case is given in the appendix.

To evaluate Eq.~\eqref{W3}, we must decompose $U(\vec k, t) = \sum_{\nu=1}^n e^{-\ii \epsilon^\nu} |\vec s^\nu\rangle \langle \vec s^\nu|$
into bands $\nu = 1, \dots, n$ with quasi\-energies $\epsilon^\nu \equiv \epsilon^\nu(\vec k,t)$ and eigenvectors $\vec s^\nu \equiv \vec s^\nu(\vec k, t)$.
Quasienergies are measured in units of $1/T$, and defined up to multiples of $2 \pi$.
We assume that the $\epsilon^\nu(\vec k,t)$ are continuous functions.
At $t=T$, $\epsilon^\nu(\vec k, T)$ agrees (modulo $2 \pi$) with the Floquet quasienergy  derived from the eigenvalues of $U(\vec k, T)$.

A degeneracy point  $\vec d_i = (\vec k_i, t_i, \epsilon_i)$ occurs whenever
the quasienergies $\epsilon^{\nu}(\vec k_i,t_i)$, $\epsilon^{\mu}(\vec k_i,t_i)$ 
 of two bands $\nu \ne \mu$ 
differ by a multiple of $2 \pi$,
such that 
$ e^{-\ii \epsilon^\nu} = e^{-\ii \epsilon^{\mu}} = e^{-\ii \epsilon_i} $
for two eigenvalues of $U(\vec k_i, t_i)$.
With each degeneracy point, we can associate the Chern numbers $C^{\nu}(\vec d_i) =\oint_{\mathcal S(\vec d_i)} F_{\alpha}^{\nu} \; \mathrm dS^{\alpha}$, given as the integral of the Berry curvature $2 \pi \ii F_\alpha^\nu(\vec k,t)=\epsilon_{\alpha \beta \gamma}\partial^{\beta}\big(\vec s^{\nu}(\vec k,t)^\dagger \, \partial^{\gamma} \vec s^{\nu}(\vec k,t)\big)$ over a small surface $\mathcal S(\vec d_i)$ enclosing the degeneracy point.
It is $C^{\nu}(\vec d_i) = - C^{\mu}(\vec d_i) \ne 0$ only for the bands $\nu, \mu$ that touch in the degeneracy point.

The contribution from each degeneracy point is multiplied by the integer
$N^\nu(\epsilon, \vec d_i) = \lceil (\epsilon^\nu(\vec k_i, t_i) - \epsilon)/(2 \pi) \rceil + \lceil (\epsilon-\epsilon^\nu(\vec k, T))/(2 \pi) \rceil$
 that counts how often band $\nu$ crosses the gap at $\epsilon$ while it evolves from the degeneracy point at $t = t_i$ to its final position at $t=T$.
Here, $\lceil \cdot \rceil$ denotes rounding up to the next integer. 
Since $\epsilon$ lies in a gap, $N^\nu(\epsilon, \vec d_i)$ does not depend on $\vec k$.

Moving from one gap at $\epsilon$ to the next gap at $\epsilon'$, both being separated by band $\nu$, the value of $N^\nu(\epsilon, \vec d_i)$ changes by one, such that $W_3(\epsilon)$ changes by $C^\nu = \sum_{i=1}^{\mathrm{dp}} C_i^{\nu}(\vec d_i)$.
The value of $C^\nu$ is just the Chern number of band $\nu$ at $t=T$. 
Note that when we move once through the quasienergy spectrum, letting $\epsilon \mapsto \epsilon + 2 \pi$, we change $W_3(\epsilon)$ by $\sum_\nu C^\nu = 0$.

\begin{figure}
\hspace*{\fill}
\includegraphics[width=0.95\linewidth]{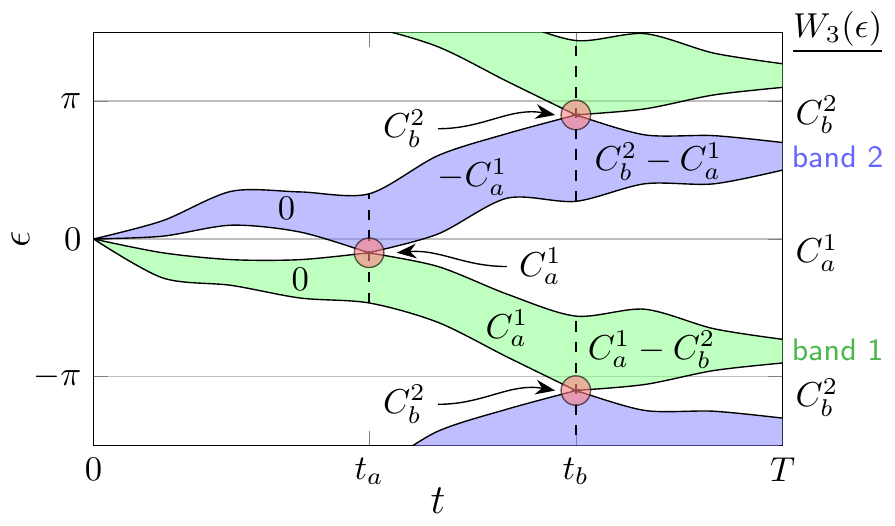}
\hspace*{\fill}
\caption{Schematic illustration of two Floquet-Bloch bands $\nu = 1,2$, which touch in two degeneracy points $i=a,b$ during the time evolution from $t=0$ to $t=T$.
At each degeneracy point, the Chern numbers of the bands and the $W_3$-invariant change by the integer $C_i^1(\vec d_i) = - C_i^2(\vec d_i)$ according to Eq.~\eqref{W3}.
In the situation sketched here, anomalous boundary states occur if $C_a^1 = C_b^2 \ne 0$,
such that the bands are topologically trivial at $t=T$ but $W_3 \ne 0$ in each gap.
}
\label{fig:Sketch}
\end{figure}

In the situation sketched in Fig.~\ref{fig:Sketch}, 
we have $N^\nu(\epsilon, \vec d_i) = 1$ (or $N^\nu(\epsilon, \vec d_i) = 0$)
for the band directly below (or above) the gap at $\epsilon$.
Here, where the bands of $U(\vec k, t)$ do not wind around the circle independently, $W_3(\epsilon)$ is simply the sum over the degeneracy points in each gap.

\section{$\mathbb Z_2$-invariants for Floquet-Bloch systems with symmetries}
\label{sec:Z2}

For the construction of the new $\mathbb Z_2$-invariants we adapt Eq.~\eqref{W3}, essentially by including only half of the degeneracy points in the summation.
We will now, for each of the three symmetries, introduce the respective invariant,
formulate the bulk-boundary correspondence between the invariant and symmetry-protected topological boundary states,
and present an exemplary application to a 
Floquet-Bloch system with the specific symmetry.

\subsection{Chiral symmetry.}

\begin{figure*}
\hspace*{\fill}
\raisebox{25pt}{\includegraphics[scale=1]{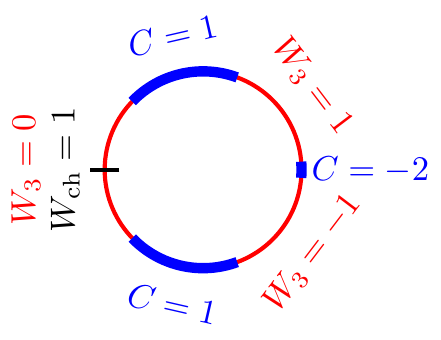}}
\hspace*{\fill}
\includegraphics[scale=0.3]{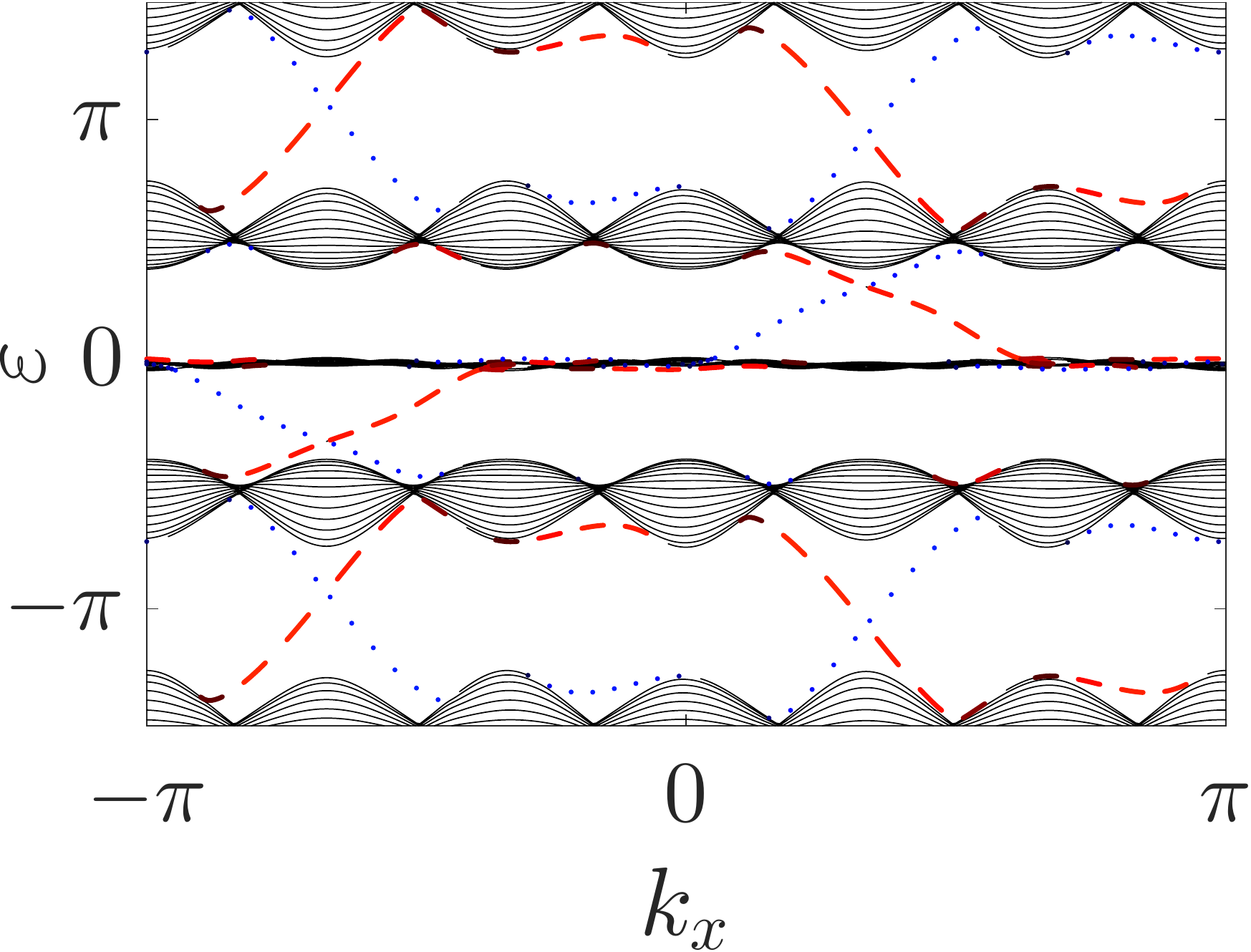}
\hspace*{\fill}
\includegraphics[scale=0.3]{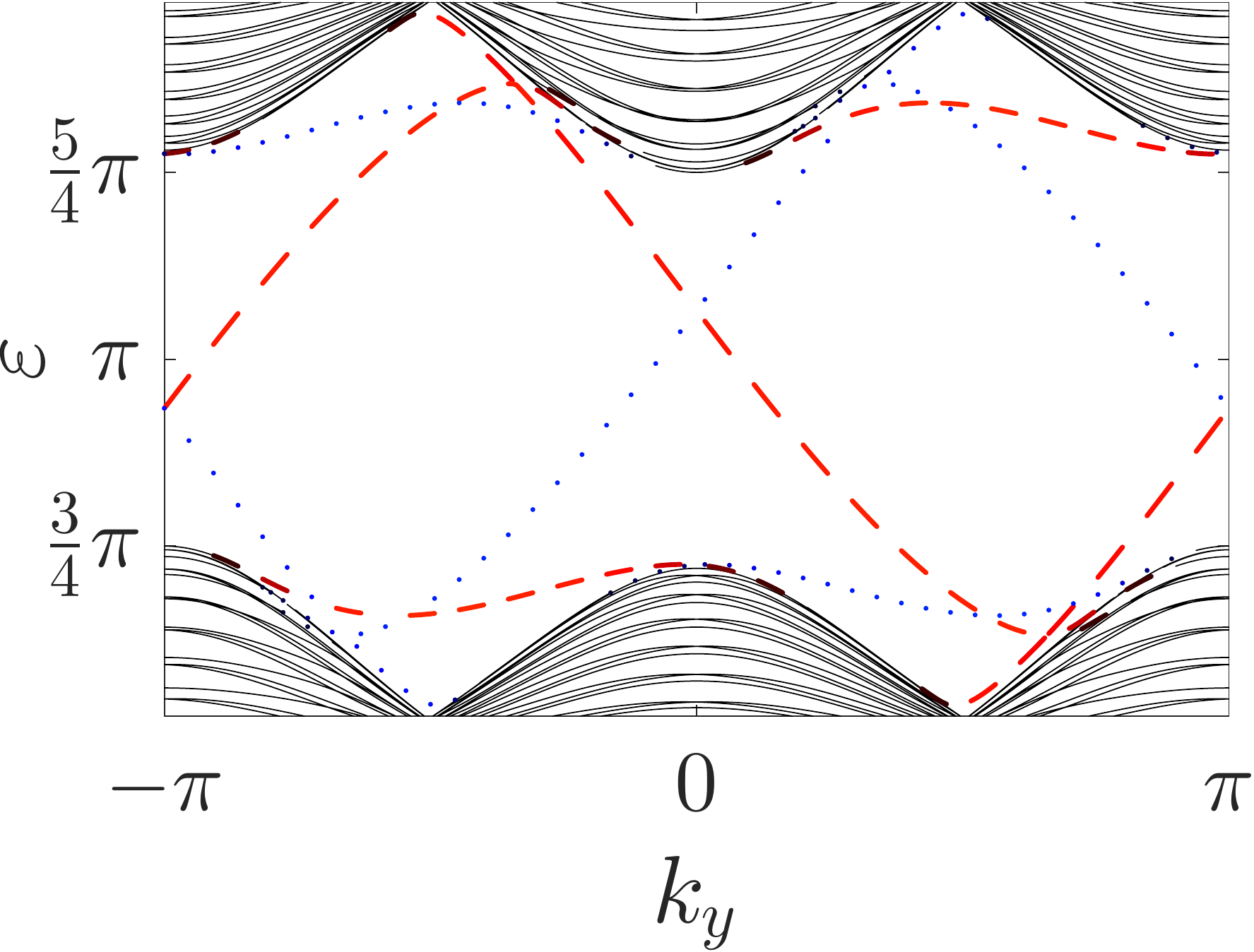}
\hspace*{\fill}
\caption{Bands and boundary states for the chiral model~\eqref{CH_Ham} at $t=T$.
Left panel: Diagrammatic representation of the Floquet bands $\exp(-\ii \epsilon^\nu(\vec k, T))$ on the circle $\mathbb S^1$ (indicated by thick arcs),
with three gaps at quasienergies $\epsilon = \pm \pi/4$ and $\epsilon = \pi$.
Included are the Chern numbers of each band,
and the $W_3$ and $W_\mathrm{ch}$-invariants in each gap.
Central and right panel:  Bands (solid) and boundary states (dashed/dotted) on a semi-infinite strip along the $x$ or $y$-axis,
as a function of momentum $k_x$ or $k_y$. Dashed red (dotted blue) curves correspond to boundary states on the top (bottom) boundary. In the right panel, we show only the gap at $\epsilon=\pi$ for better visibility. 
For both boundary orientations, one pair of symmetry-protected topological boundary states exists in the gap at $\epsilon=\pi$ in accordance with $W_\mathrm{ch}(\pi) \ne 0$ in the left panel.
}
\label{fig:CH}
\end{figure*}

The symmetry relation for chiral symmetry, realized as a sublattice symmetry on a bipartite lattice, is
\begin{equation}\label{CH_Symm}
 H_\mathrm{ch}(\vec k, t) = - S H_\mathrm{ch}(\vec k + \vec k_\pi ,T-t) S^{-1}
\end{equation}
with a unitary operator $S$, and a reciprocal lattice vector $\vec k_\pi$ corresponding to the sublattice decomposition (e.g., $\vec k_\pi = (\pi,\pi)$ for a square lattice).

Note that the symmetry relation~\eqref{CH_Symm} differs from the standard definition of chiral symmetry~\cite{PhysRevB.93.115429, PhysRevB.96.155118,PhysRevB.96.195303},
which does not contain the momentum shift $\vec k_\pi$. 
The inclusion of the momentum shift $\vec k_\pi$ is crucial for the existence of symmetry-protected boundary states and of the $\mathbb Z_2$-invariant defined below.
As detailed in App.~\ref{app:Ch},
a Hamiltonian $H_\mathrm{ch}(\cdot)$ that fulfills Eq.~\eqref{CH_Symm} also fulfills the standard chiral symmetry relation but possesses an additional symmetry that protects the topological phases and boundary states.
Without the $\vec k_\pi$--shift, chiral symmetry does not allow for the symmetry-protected boundary states observed here~\cite{PhysRevB.93.115429, PhysRevB.96.155118,PhysRevB.96.195303}.

Because of the $T-t$ argument on the right hand side, the symmetry relation~\eqref{CH_Symm} 
does not extend to $U(\vec k , t)$ but only to the time-symmetrized propagator
$U_{\star}(\vec k,t) = U(\vec k, \tfrac12 (t + T) ) \, U^\dagger(\vec k, \tfrac12 (T-t))$,
for which it implies $SU_{\star}(\vec k + \vec k_\pi,t)S^{-1} = U_{\star}^\dagger(\vec k, t)$.
Therefore, degeneracy points of $U_\star(\cdot)$ occur in pairs $\vec d_i = (\vec k_i, t_i, \epsilon_i)$,
$\hat{ \vec d}_i = (\vec k_i + \vec k_\pi, t_i, -\epsilon_i)$ with opposite sign of $C^\nu(\vec d_i) = - C^\nu(\hat{\vec d}_i)$.
The $W_3$-invariant, computed from $U_\star(\cdot)$, fulfills $W_3(-\epsilon) = - W_3(\epsilon)$, especially $W_3(\epsilon) = 0$ for a gap at $\epsilon = 0, \pi$. 

Note that $U_\star(\cdot)$ belongs to a family of propagators that are related to $U(\cdot)$ by the homotopy
$s \mapsto U\left(\vec k,(1-s) t+sT\right)U^{\dagger}\left(\vec k,s(T-t)\right)$.
For $s=0$, we obtain the original propagator $U(\cdot)$, for $s=1/2$ the symmetrized propagator $U_\star(\cdot)$.
Since $U_\star(\cdot)$ is homotopic to $U(\cdot)$, with fixed boundary values $U_{\star}(\vec k, 0) = 1$ and $U_{\star}(\vec k, T) = U(\vec k, T)$, we obtain the same result if $W_3(\epsilon)$ is computed with the original propagator $U(\cdot)$.
In this computation, however, the cancellation of degeneracy points would not be obvious.

We now define a $\mathbb Z_2$-invariant, for $\epsilon = 0$ or $\epsilon=\pi$, via
\begin{equation}\label{W_CH}
W_{\mathrm{ch}}(\epsilon) \equiv \sum_{\nu=1}^n \sum_{i=1}^{\mathrm{dp}/2} N^{\nu}(\epsilon, \vec d_i) \, C^{\nu}(\vec d_i) \mod 2 \; ,
\end{equation}
where the upper limit $\mathrm{dp}/2$ in the sum over $i$ indicates that exactly one degeneracy point of each symmetric pair 
$\vec d_i$, $\hat{ \vec d}_i$ is included.
Depending on which points are included the sum can differ by an even number,
such that $W_{\mathrm{ch}}(\epsilon) \in \mathbb Z_2$.
Since the degeneracy points in each pair are separated by $\vec k_\pi$,
a homotopy of $H_\mathrm{ch}(\cdot)$ that respects chiral symmetry cannot annihilate the degeneracy points.
Therefore, $W_{\mathrm{ch}}(\epsilon)$ is invariant under such a homotopy.

A non-zero value of $W_\mathrm{ch}(\epsilon)$ indicates that 
an odd number of pairs of degeneracy points occur in the gap at $\epsilon$ during time-evolution from $0$ to $T$.
If a boundary is introduced into the system, say along the $x$-direction,
the first pair of degeneracy points $\vec d_i$, $\hat{ \vec d}_i$ gives rise to two boundary states $\mathrm B_\mathrm {I}$, $\mathrm B_\mathrm {II}$ of opposite chirality that appear immediately after $t_i$
at momenta $(\vec k_i)_x$, $(\vec k_i + \vec k_\pi)_x$.
During the subsequent time-evolution the dispersion of these boundary states is related by $\epsilon_\mathrm{I}(k_x) = - \epsilon_\mathrm{II}(k_x + \pi)$ due to chiral symmetry.
Therefore, the boundary states are protected:
They cannot annihilate each other, because the number of crossing through $\epsilon=0, \pi$ is fixed by the above relation.
The pair of boundary states can disappear only through the appearance of a second pair of degeneracy points at a later $t _j > t_i$.
In this way, each pair flips the value of $W_\mathrm{ch}(\epsilon)$ and the number of symmetry-protected boundary states in the respective gap.
This consideration establishes the bulk-boundary correspondence for chiral symmetry:
A non-zero bulk invariant $W_\mathrm{ch}(\epsilon)$ corresponds to the existence of a pair of symmetry-protected boundary states with opposite chirality in the gap at $\epsilon$.

\begin{figure*}
\hspace*{\fill}
\raisebox{25pt}{\includegraphics[scale=1]{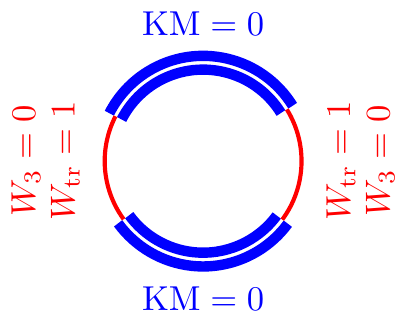}}
\hspace*{\fill}
\includegraphics[scale=0.3]{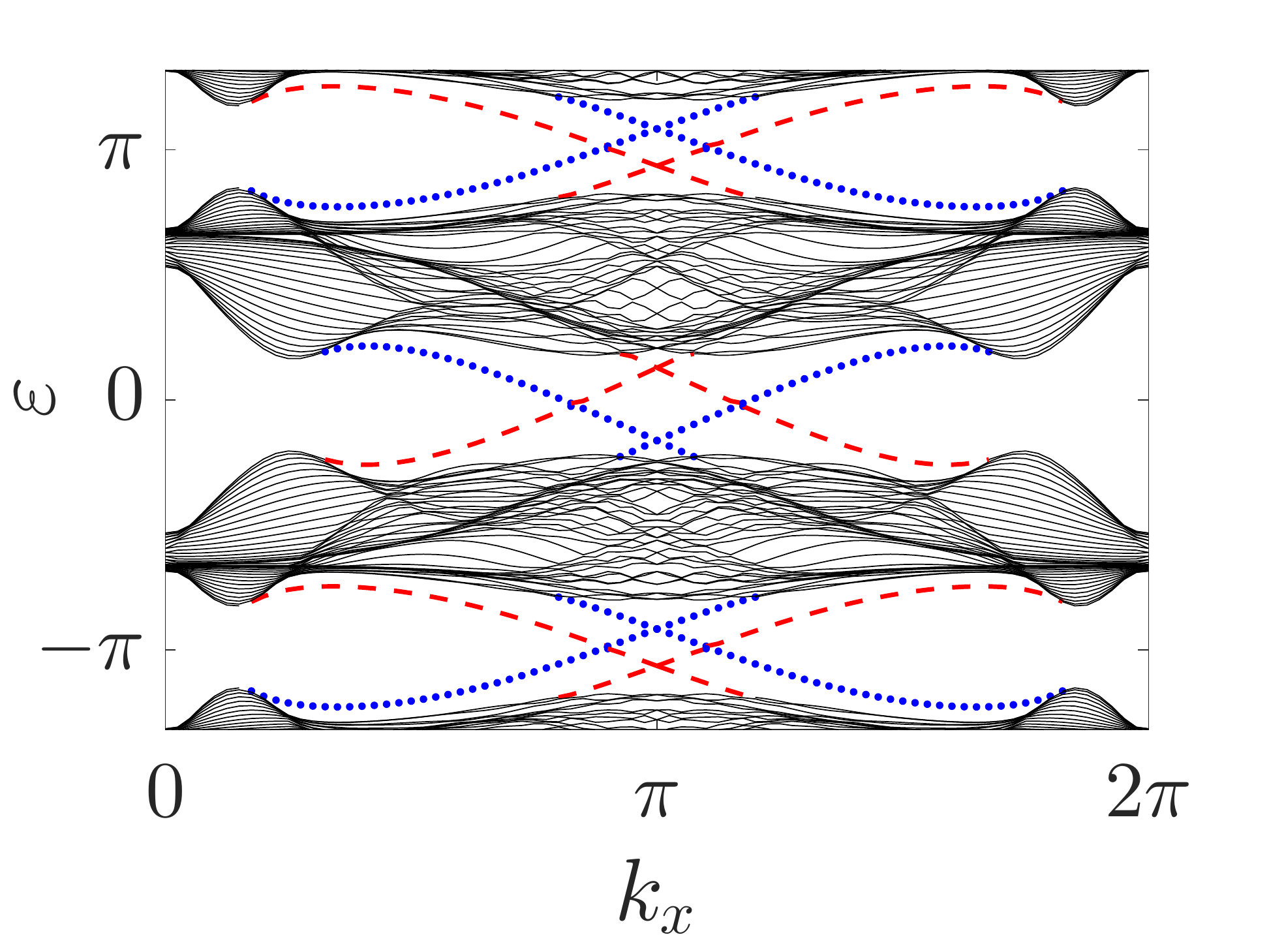}
\hspace*{\fill}
\includegraphics[scale=0.3]{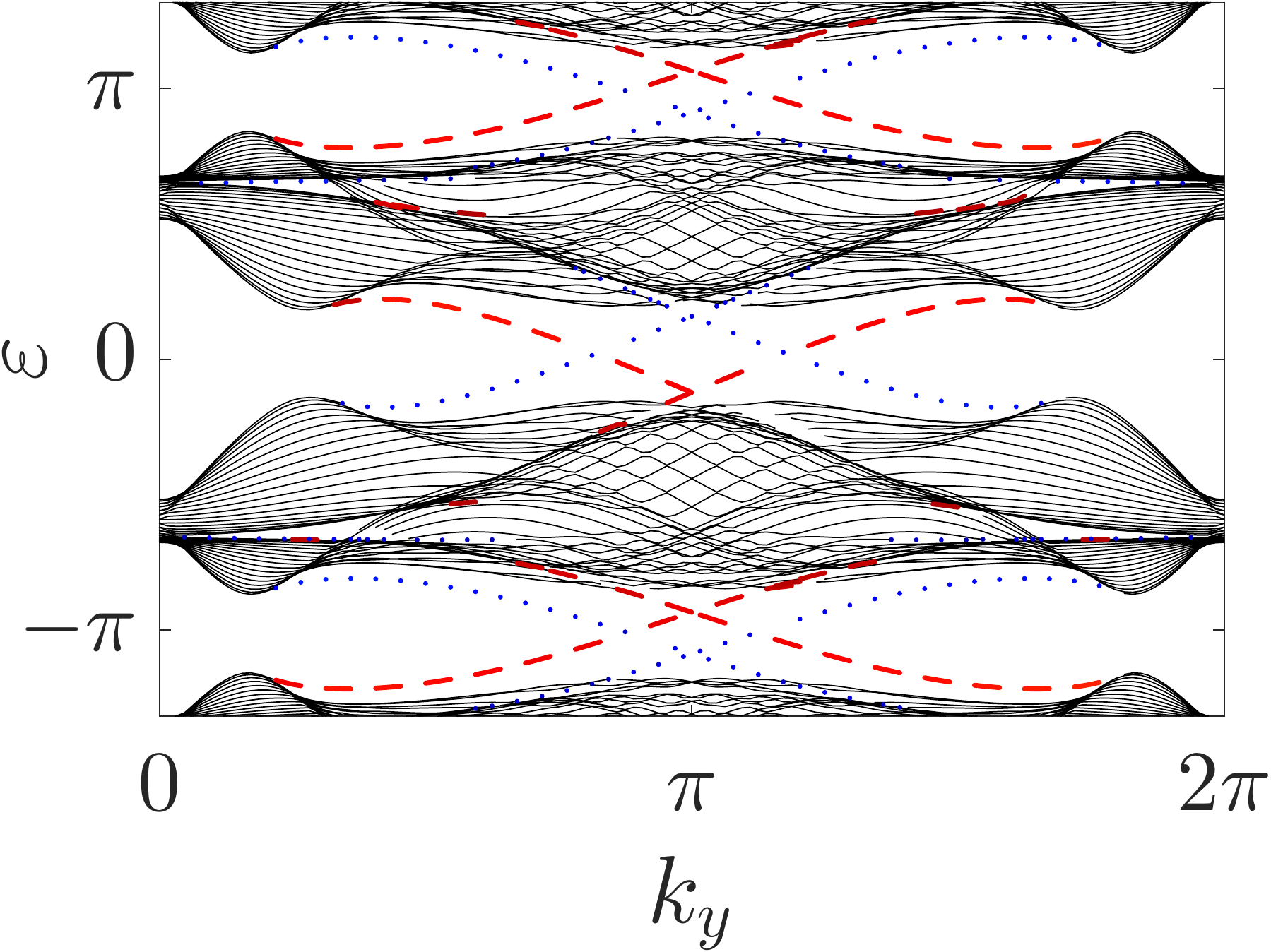}
\hspace*{\fill}
\caption{Same as Fig.~\ref{fig:CH}, now for the time-reversal model~\eqref{TR_Ham}.
Left panel: Included are the Kane-Mele invariants (KM) of each Kramers pair, 
and the $W_3$ and $W_\mathrm{tr}$-invariants in the two gaps at $\epsilon = 0, \pi$.
Central and right panel:  Bands and boundary states on a semi-infinite strip along the $x$-axis and $y$-axis.
For both boundary configurations, one pair of symmetry-protected topological boundary states exists in the two gaps in accordance with $W_\mathrm{tr}(\epsilon) \ne 0$ in the left panel.
}
\label{fig:TR}
\end{figure*}

Chiral symmetry is realized in the extended Harper model on a square lattice \cite{PhysRevB.90.205108}
\begin{multline}\label{CH_Ham}
 H_\mathrm{ch}(t) = \sum_{ij} \Big[ J_x(t) \big( e^{2\pi \ii \alpha j} c^\dagger_{i+1,j} c_{ij} + h.c. \big) + \\ J_y \big( c^\dagger_{i,j+1} c_{ij} + h.c. \big)  \Big] \;,
 \end{multline}
provided that $J_x(T-t) = J_x(t)$. The rational parameter $\alpha=p/n$ controls the number $n$ of Floquet bands.
Note that the (magnetic) unit cell of this model has one element in $x$-direction and $n$ elements in $y$-direction.

For the results in Fig.~\ref{fig:CH} we set $J_x(t)=J_{x,1}+J_{x,2} \cos(2 \pi t/T)$, with
$\alpha=1/3, J_{x,1}=2, J_{x,2}=1, J_y=2$. Since $n$ is odd, chiral symmetry prevents  the opening of a gap at $\epsilon =0$. In the gap at $\epsilon = \pi$,
where $W_3(\pi)=0$, a pair of symmetry-protected boundary states exists in accordance with the non-zero value of $W_\mathrm{ch}(\pi)$.
Note for the interpretation of Fig.~\ref{fig:CH} that according to the magnetic unit cell for $\alpha=1/3$ the two boundary states along the $y$-axis can coexist at three different quasienergies for a given $k_y$, but indeed cross the gap at $\epsilon=\pi$ only once with opposite chirality.

In summary, we see that 
Eq.~\eqref{W_CH} defines a $\mathbb Z_2$-valued bulk invariant for chiral symmetry,
which predicts the appearance (or absence) of a symmetry-protected topological phase and of the corresponding boundary states.
A different $\mathbb Z_2$-invariant,
which is constructed for a finite system with absorbing boundaries, has been introduced in Ref.~\onlinecite{PhysRevB.93.075405},
where also the `weak' or `strong' nature of topological phases with chiral symmetry is addressed.
To relate these results to our $\mathbb Z_2$-invariant we include in App.~\ref{app:Ch} additional data for different boundary orientations in the Harper model from Eq.~\eqref{CH_Ham}.

\subsection{Time-reversal symmetry.}

The symmetry relation for time-reversal symmetry of fermionic particles is
\begin{equation}\label{TR_symm}
 H_\mathrm{tr}(\vec k, t) = \Theta H_\mathrm{tr}(-\vec k ,T-t) \Theta^{-1} \;,
\end{equation}
with an anti-unitary operator $\Theta$ for which $\Theta^2=-1$.
The symmetry relation~\eqref{TR_symm} implies
$\Theta U_\star(-\vec k,t) \Theta^{-1} = U_\star^{\dagger}(\vec k,t)$, again for the time-symmetrized propagator $U_\star(\cdot)$. Therefore, degeneracy points of $U_\star(\vec k,t)$ occur in pairs $\vec d_i = (\vec k_i,t_i, \epsilon_i)$, $\hat{\vec d}_i = (-\vec k_i, t_i, \epsilon_i)$ with opposite sign of $C^\nu(\vec d_i) = - C^\nu(\hat{\vec d}_i)$. It is $W_3(\epsilon) = 0$ in each gap.

We now define a $\mathbb Z_2$-invariant
\begin{equation}\label{W_TR}
W_{\mathrm{tr}}(\epsilon) \equiv \sum_{\nu=1}^{2n} \sum_{i=1}^{\mathrm{dp}/2} N^{\nu}(\epsilon, \vec d_i) \, C^{\nu}(\vec d_i) \mod 2 \; ,
\end{equation}
where again only one degeneracy point from each symmetric pair is included in the sum.

Note that the bands of $U_\star(\cdot)$ appear in Kramers pairs~\cite{PhysRevLett.95.146802} which, if arranged in this specific order, fulfill
$\epsilon^{2\nu-1}(-\vec k,t)=\epsilon^{2\nu}(\vec k,t)$. 
The two bands of each Kramers pair are degenerate at the invariant momenta (IM) $\vec k \equiv - \vec k$
(modulo a reciprocal lattice vector).
The Kramers degeneracy at the IM, which is 
enforced by time-reversal symmetry for all $t$,
must be distinguished from the degeneracy points that contribute in Eq.~\eqref{W_TR}:
These occur only at certain $t_i$ and involve two bands from two different Kramers pairs.

The considerations leading to a bulk-boundary correspondence are similar to those for chiral symmetry.
Again, a pair of degeneracy points $\vec d_i$, $\hat{\vec d}_i$ gives rise to two boundary states, which
now appear at momenta $(\vec k_i)_x$, $- (\vec k_i)_x$.
Their dispersion relations are connected by $\epsilon_\mathrm{I}(k_x) = \epsilon_\mathrm{II}(-k_x)$,
with Kramers degeneracy at the IM $k_x \equiv - k_x$.
Because of $\Theta^2 = -1$ the boundary states are two-fold degenerate at the IM,
which prevents their mutual annihilation.
Continuing with the reasoning as before, we conclude that a non-zero value of $W_{\mathrm{tr}}(\epsilon)$ implies the existence of a pair of symmetry-protected boundary states with opposite chirality in the gap at $\epsilon$. 

If we move from one gap at $\epsilon$ to the next gap at $\epsilon'$, separated by a Kramers pair of bands $2 \nu -1$, $2 \nu$, the value of $W_\mathrm{tr}(\epsilon)$ changes by $W_\mathrm{tr}(\epsilon') -  W_\mathrm{tr}(\epsilon) \equiv  \sum_{i=1}^{\mathrm{dp}/2} ( C^{2 \nu-1}(\vec d_i) + C^{2 \nu}(\vec d_i)) \mod 2$.
The right hand side of this expression gives just the Kane-Mele invariant~\cite{PhysRevLett.95.146802} of the respective Kramers pair (see App.~\ref{app:KM}).

Time-reversal symmetry is realized in the extended Kane-Mele model on a graphene lattice \cite{driven_KM}
\begin{equation}\label{TR_Ham}
\begin{split}
 H_\mathrm{tr}(t) = J_1(t) \sum_{\langle i,j \rangle} c^\dagger_i c_j \, + \, \ii J_2(t) \sum_{\langle\langle i,j \rangle\rangle} \nu_{ij} c^\dagger_i \sigma_z c_j \\
 + \,
 \lambda_\nu \sum_i \xi_i c^\dagger_i c_i + \ii \lambda_R \sum_{\langle i,j \rangle} c^\dagger_i (\boldsymbol \sigma \times \boldsymbol d_{ij})_z  c_j
  \;,
 \end{split}
\end{equation}
provided that 
$J_{1,2}(T-t) = J_{1,2}(t)$.
For the results in Fig.~\ref{fig:TR} we set 
$J_1(t)=J_a+J_b \cos(2 \pi t/T)$, $J_2(t)=J_c+J_d \cos(2 \pi t/T)$
with $J_a=0.9, J_b=1.8, J_c=0.6, J_d=1.2$, and $\lambda_\nu=1.8, \lambda_{\mathrm R}=0.3$.  The $W_{\mathrm{tr}}$-invariant correctly predicts the appearance of symmetry protected boundary states in the gaps at $\epsilon=0$ and $\epsilon=\pi$, while the Kane-Mele invariants of the Floquet bands and the $W_3$-invariant vanish.

\begin{figure*}
\hspace*{\fill}
\raisebox{25pt}{\includegraphics[scale=1]{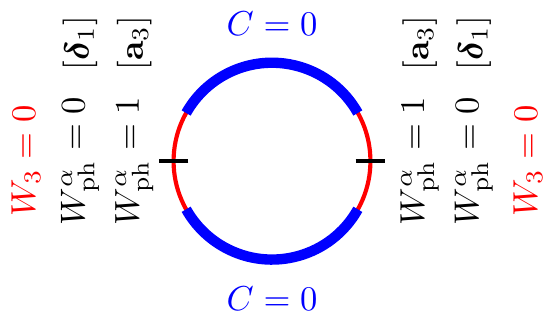}}
\hspace*{\fill}
\includegraphics[scale=0.3]{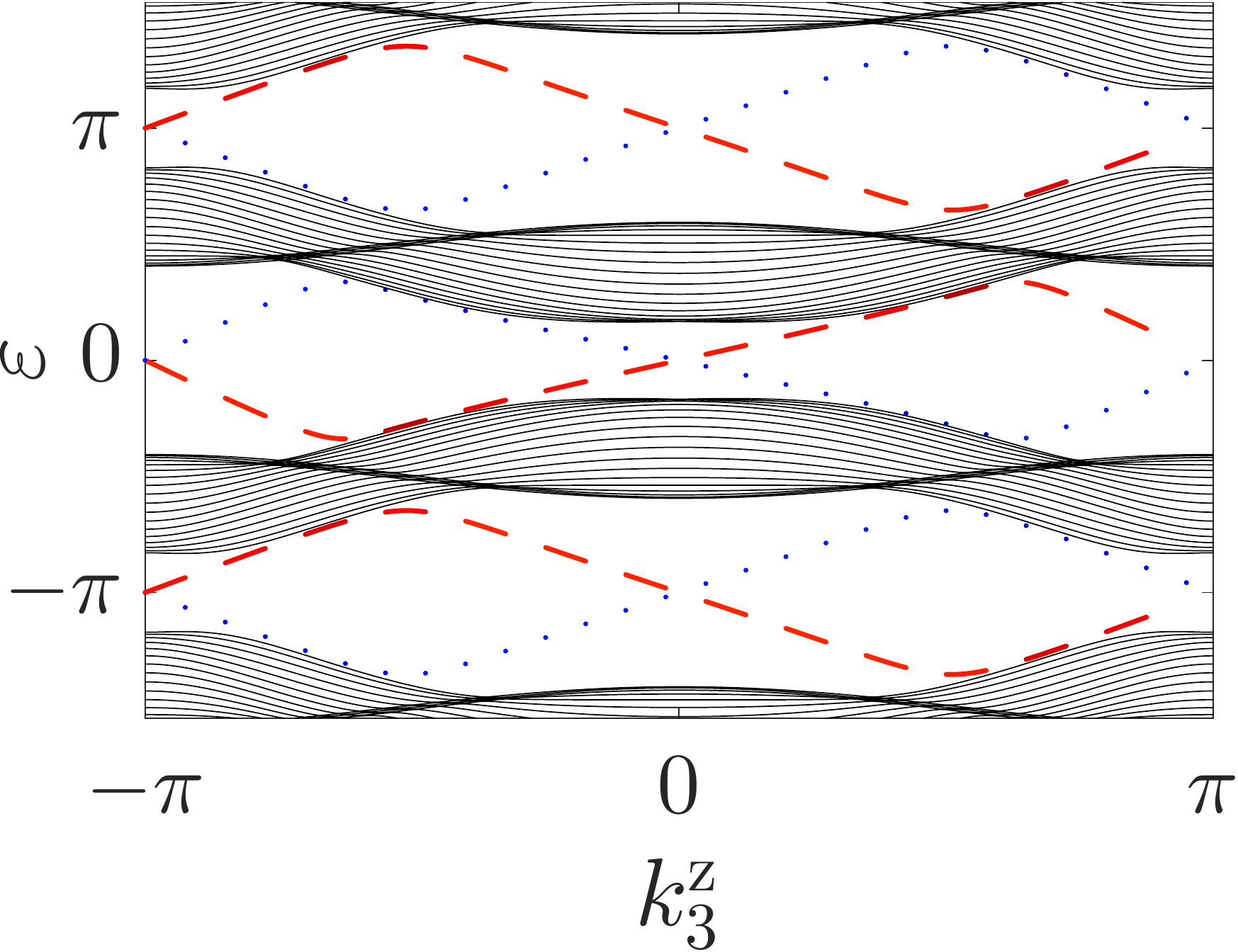}
\hspace*{\fill}
\includegraphics[scale=0.3]{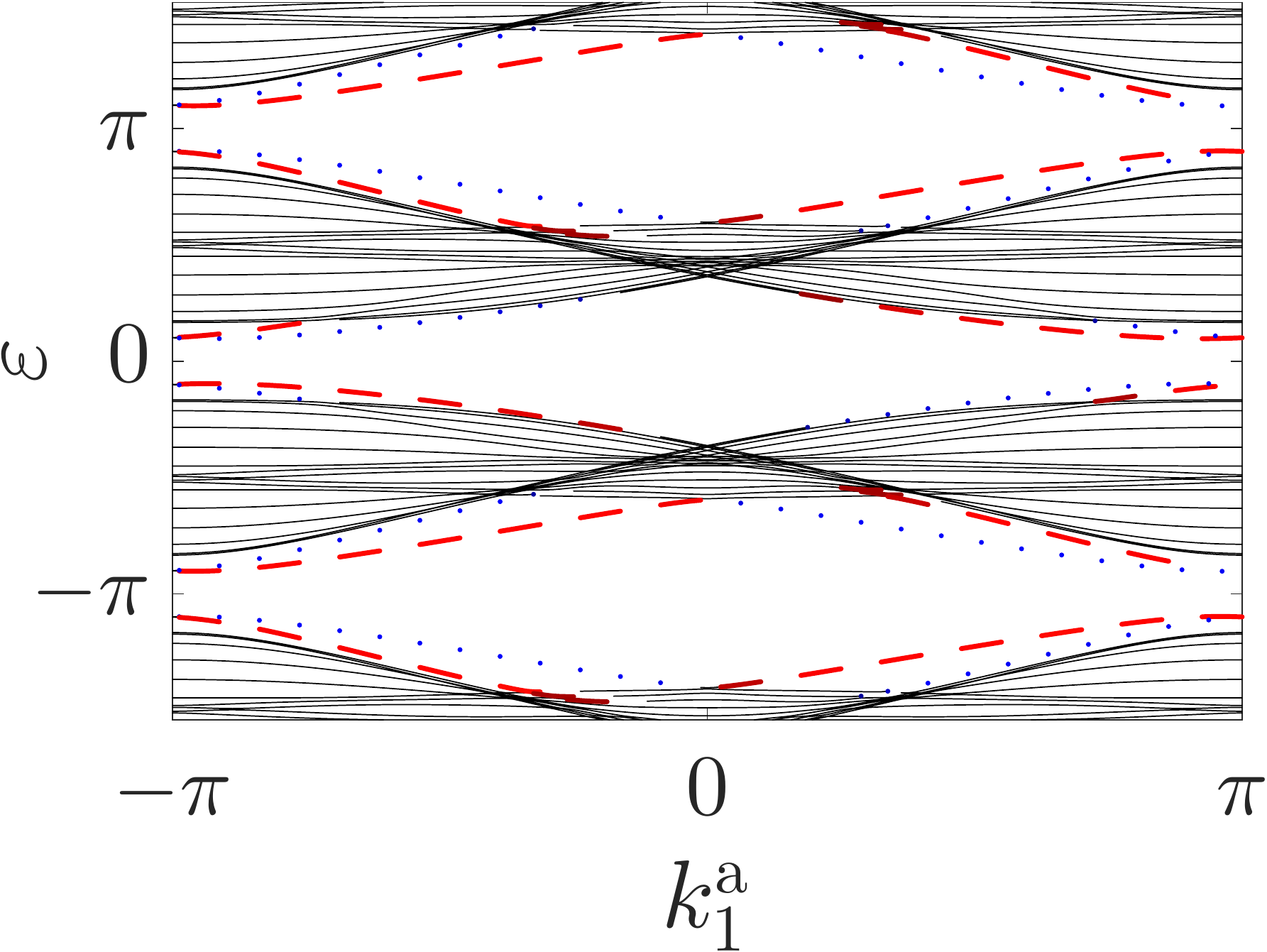}
\hspace*{\fill}
\caption{Same as Figs.~\ref{fig:CH},~\ref{fig:TR}, now for the particle-hole symmetric model~\eqref{PH_Ham}. 
Left panel: Included are the Chern numbers of each band,
and the $W_3$ and $W_\mathrm{ph}$-invariants in the two gaps at $\epsilon = 0, \pi$.
Central (right) panel:  Bands and boundary states on a semi-infinite strip with boundaries along the $\vec a_3$ (or $\boldsymbol \delta_1$) direction, as a function of the respective momentum $k_3^z$ (or $k_1^a$) parallel to the zigzag (or armchair) boundary.
In both gaps, symmetry-protected boundary states exist at $k_3^z = 0, \pi$ (or are absent at $k_1^a = 0, \pi$) in accordance with
$W^{0}_\mathrm{ph} = W^{\pi}_\mathrm{ph} \ne 0$ for $\vec a_3$
(or $W^{0}_\mathrm{ph} = W^{\pi}_\mathrm{ph} = 0$ for $\boldsymbol \delta_1$)
 in the left panel.}
\label{fig:PH}
\end{figure*}

In summary, we see that Eq.~\eqref{W_TR} defines a $\mathbb Z_2$-valued bulk invariant for time-reversal symmetry.
The construction of this invariant closely resembles the construction from Ref.~\onlinecite{1367-2630-17-12-125014},
to which it reduces under the additional conditions stated in App.~\ref{app:W3} for the $W_3$-invariant.
A different $\mathbb Z_2$-invariant has been introduced in Ref.~\onlinecite{PhysRevLett.114.106806},
which is based on the original expression~\cite{PhysRevX.3.031005} for the $W_3$-invariant and requires a 
more complicated auxiliary construction~\cite{PhysRevLett.114.106806, CARPENTIER2015779} of a time-symmetrized propagator.

\subsection{Particle-hole symmetry.}

The symmetry relation for particle-hole symmetry of fermionic particles is
\begin{equation}\label{PH_Symm}
 H_\mathrm{ph}(\vec k, t) = - \Pi H_\mathrm{ph}(-\vec k ,t) \Pi^{-1} \;,
\end{equation}
with an anti-unitary operator $\Pi$ for which $\Pi^2=1$.
The symmetry relation~\eqref{PH_Symm} implies $\Pi U(-\vec k,t) \Pi^{-1}= U(\vec k,t)$, for the original propagator $U(\cdot)$. If degeneracy points of $U(\vec k,t)$ occur in pairs $\vec d_i = (\vec k_i,t_i, \epsilon_i)$, $\hat{\vec d}_i = (-\vec k_i,t_i, -\epsilon_i)$, they now occur with the same sign of $C^\nu(\vec d_i) = C^\nu(\hat{\vec d}_i)$.
We can only conclude $W_3(\epsilon) = W_3(-\epsilon)$,
and in contrast to chiral and time-reversal symmetry
the symmetry relation does not enforce $W_3(\epsilon)=0$ in any gap.

Despite this difference, symmetry-protected boundary states exist also for particle-hole symmetry, because
the IM $\vec k \equiv - \vec k$ again have specific significance
but play the opposite role as in the case of time-reversal symmetry.
There, $\Theta^2 = -1$ forbids single unpaired boundary states at the IM,
while here $\Pi^2 = 1$ is compatible with their appearance.
An unpaired boundary state in the gaps at $\epsilon = 0,\pi$, which is pinned at the IM, is protected by particle-hole symmetry~\cite{PhysRevB.89.104523,PhysRevB.93.075405}.
These states are associated with unpaired degeneracy points of $U(\vec k, t)$ at the IM.

Let the four IM in the $2+1$ dimensional bulk system be  
$\vec M_{0} = 0$,
$\vec M_{1} = \vec b_1/2$,
$\vec M_{2} = \vec b_2/2$,
$\vec M_{3} = (\vec b_1 + \vec b_2)/2$,
for two primitive reciprocal lattice vectors $\vec b_1$, $\vec b_2$.
If we introduce a boundary 
along a primitive lattice vector $\vec a$,
with $\vec a \cdot \vec b_{1,2} \in \{ 0, 2\pi \}$,
the four IM are projected onto two momenta $k_{\vec a} = \vec a \cdot \vec M_m \in \{ 0, \pi \}$.
Symmetry-protected boundary states,
with dispersion relation $\epsilon(-k_{\vec a}) = - \epsilon(k_{\vec a})$,
can exist at both momenta.

To capture this situation, we need a total of four $\mathbb Z_2$-invariants, defined for $\alpha =0, \pi$ and $\epsilon = 0, \pi$ as
\begin{equation}\label{W_PH}\!\!
\;\;\;\;  W^{\alpha}_\mathrm{ph}(\epsilon) =  \sum_{\nu=1}^n \; \sum_{\substack{\vec k_i \in \{ \vec M_m\} \\ \vec a \cdot \vec k_i = \alpha} } \!\!\!\!\! N^{\nu}(\epsilon, \vec d_i) \, C^{\nu}(\vec d_i) \!\! \mod 2 \;.
\end{equation}
In Eq.~\eqref{W_PH} only unpaired degeneracy points at the two IM $\vec M_m$ with $\vec a \cdot \vec M_m = \alpha$ contribute.
Therefore, a non-zero $W_\mathrm{ph}$-invariant implies the existence of a symmetry-protected boundary state that is pinned at the respective momentum $k_{\vec a} = \alpha$.
For example, $W^{\pi}_\mathrm{ph}(0) \ne 0$ corresponds to a symmetry-protected boundary state with $\epsilon(k_{\vec a} = \pi) = 0 $ in the gap at $\epsilon=0$.
Note that we assume here the absence of boundary states for $t = 0$ (but see App.~\ref{app:PH} for an extended discussion).

The $W_\mathrm{ph}$-invariants only count unpaired degeneracy points,
which necessarily occur at IM.
The $W_3$-invariant also counts paired degeneracy points with opposite momenta $\pm \vec k_i$ that occur away from the IM.
Since paired degeneracy points change the $W_3$-invariant by an even number, we have $W^{0}_\mathrm{ph}(\epsilon) + W^{\pi}_\mathrm{ph}(\epsilon) \equiv W_3(\epsilon) \mod 2$.

According to the summation in Eq.~\eqref{W_PH} the $W_\mathrm{ph}$-invariants depend on the boundary orientation given by $\vec a$.
Especially if $W_3(\epsilon) = 0$ a `weak' topological phase can occur~\cite{PhysRevB.89.104523},
where two symmetry-protected boundary states exist on some boundaries where $W^0_\mathrm{ph} = W^\pi_\mathrm{ph} = 1$,
but not on other boundaries where $W^0_\mathrm{ph} = W^\pi_\mathrm{ph} = 0$.
If, on the other hand, $W_3(\epsilon) \ne 0$ in a `strong' topological phase, boundary states occur on each boundary.
Especially for odd $W_3(\epsilon)$, we must have non-zero $W_\mathrm{ph}$ invariants for each boundary orientation,
and thus a symmetry-protected boundary state at either $k_{\vec a} = 0$ or $k_{\vec a} = \pi$.

Particle-hole symmetry is realized in the graphene lattice model ~\cite{PhysRevB.82.235114,PhysRevB.93.075405}
\begin{equation}\label{PH_Ham}
 H_\mathrm{ph}(t) = \sum_{\vec r} \sum_{l=1}^3 \, J_l(t) \, c_{B,\vec r}^\dagger c_{A,\vec r + \boldsymbol \delta_l} + h.c. \;,
\end{equation}
without further constraints on the $J_l(t)$.
The $J_l(t)$ 
are periodically varied according to the protocol in Ref.~\onlinecite{PhysRevB.93.075405}.
For the results in Fig.~\ref{fig:PH} we set
$J_{s,1}=-3\pi/2, J_{s,2}=-3\pi/2, J_{s,3}=3\pi/2, J_{u,1}=0, J_{u,2}=-1.2, J_{u,3}=0.9$.

In Fig.~\ref{fig:PH}  we recognize the weak topological phase just discussed:
On a zigzag boundary along a lattice vector $\vec a_3$, with invariants $W^{0}_\mathrm{ph}(\epsilon) = W^{\pi}_\mathrm{ph}(\epsilon) \ne 0$, we observe in each gap two symmetry-protected boundary states with opposite chirality at momenta $k_3^z = 0, \pi$.
On an armchair boundary along a nearest-neighbor vector $\boldsymbol \delta_1$,
with invariants $W^{0}_\mathrm{ph}(\epsilon) = W^{\pi}_\mathrm{ph}(\epsilon) = 0$,
no boundary states cross $\epsilon=0$ or $\epsilon=\pi$.
The $W_{\mathrm{ph}}$-invariants, together with the zero $W_3$-invariant, correctly describe this situation.

Note that for a hexagonal lattice, with three inequivalent orientations for each boundary type, an exhaustive analysis is significantly more complicated than suggested by Fig.~\ref{fig:PH}. For details we refer the reader to App.~\ref{app:PH}.

In summary, we see that Eq.~\eqref{W_PH} defines four $\mathbb Z_2$-valued bulk invariants for particle-hole symmetry, which
predict the appearance of symmetry-protected boundary states at $k_{\vec a} = 0, \pi$
 in dependence on the boundary orientation.
Since non-zero $W_\mathrm{ph}$-invariants are compatible with both $W_3(\epsilon)=0$ and $W_3(\epsilon)\ne 0$,
weak and strong topological phases can be distinguished.
The possible combinations of the four invariants for fixed $W_3(\epsilon)$ are given by the summation rule stated above.
Different $\mathbb Z_2$-invariants have been introduced in Ref.~\onlinecite{PhysRevB.93.075405}, in the form of scattering invariants for finite systems.

\section{Conclusions}
\label{sec:Conc}

The $\mathbb Z_2$-invariants introduced here allow for the classification of topological phases in driven systems with chiral, time-reversal, or particle-hole symmetry.
In this way, they complement the $W_3$-invariant for driven systems without additional symmetries.
The $\mathbb Z_2$-invariants are related to previous constructions for symmetry-protected topological phases~\cite{PhysRevB.93.075405,PhysRevLett.114.106806,CARPENTIER2015779,1367-2630-17-12-125014}, but they combine two substantial aspects.
First, they are bulk invariants of driven systems, and a bulk-boundary correspondence holds for each invariant.
Second, they are given by simple and explicit expressions that involve the (time-symmetrized) Floquet-Bloch propagator, 
but require no complicated auxiliary constructions. Quite intuitively, the invariants are defined through counting of half of the degeneracy points 
that appear in symmetric pairs.
Note that the invariants depend on the entire time evolution of $U(\vec k,t)$ over one period $0 \le t \le T$, as required for driven systems with the possibility of anomalous boundary states~\cite{PhysRevB.82.235114, PhysRevX.3.031005, 1367-2630-17-12-125014, PhysRevLett.114.106806, PhysRevB.93.075405}.
Once the degeneracy points are known computation of the invariants according to Eqs.~\eqref{W_CH},~\eqref{W_TR},~\eqref{W_PH} is straightforward.
Particularly efficient computation of the $\mathbb Z_2$-invariants is possible with the algorithm from Ref.~\onlinecite{HAF17}.

These aspects should make the $\mathbb Z_2$-invariants viable tools in the analysis of driven systems with symmetries.
For the three generic models considered here, the invariants correctly predict the appearance of symmetry-protected topological boundary states, even if the static invariants and the $W_3$-invariant vanish.
Concerning the nature of these states, chiral and time-reversal symmetry are set apart from particle-hole symmetry. In the latter case, the existence of symmetry-protected states depends on the orientation of the boundary, similar to the situation for three-dimensional weak topological insulators~\cite{PhysRevLett.98.106803} or quantum Hall systems~\cite{PhysRevB.45.13488}. 
It will be interesting to study the different impact of symmetries on topological phases, and on the anomalous boundary states that are unique to driven systems, in nature. 
One way towards realization of the proper symmetries should be offered by photonic crystals~\cite{anomalous, anomalous_2}.

\begin{acknowledgments}
 This work was financed in part by Deutsche Forschungsgemeinschaft through SFB 652.
 B.H. was funded by the federal state of Mecklenburg-West Pomerania through a postgraduate scholarship within
the International Helmholtz Graduate School for Plasma Physics. 
\end{acknowledgments}

\appendix

\section{Derivation of expression Eq.~\eqref{W3} for the $W_3$--invariant}
\label{app:W3}

We here give the details of the derivation of Eq.~\eqref{W3}.
In slightly different notation, Eqs.~(3.18) and (5.4) in Ref.~\onlinecite{HAF17} yield the expression
\begin{multline}\label{W3_old}
 W_3(\epsilon) =  \frac{1}{2 \pi}\sum_{\nu=1}^n \Bigg[ \int\limits_0^T\iint\limits_{\mathcal B} ( \partial^{\alpha} F_\alpha^{\nu}(\vec k,t) ) \epsilon^{\nu}(\vec k,t) \; \mathrm dk_1 \mathrm dk_2 \mathrm dt \\
+ \iint\limits_{\mathcal B}  F_3^{\nu}( \vec k, T ) \left(\ii \log_{\epsilon} e^{-\ii \epsilon^{\nu}(\vec k,T)}- \epsilon^{\nu}(\vec k,T) \right)   \;  \mathrm dk_1 \mathrm dk_2  \\
+ \iint\limits_{\mathcal B}  F_3^{\nu}( \vec k, 0 ) \epsilon^{\nu}(\vec k,0)  \;  \mathrm dk_1 \mathrm dk_2 \Bigg] 
\end{multline}
for the $W_3$-invariant from Ref.~\onlinecite{PhysRevX.3.031005}. 
It is written as an integral of the Berry curvature
\begin{equation}\label{Berry}
F_\alpha^\nu(\vec k,t)= \frac{1}{2 \pi \ii} \epsilon_{\alpha \beta \gamma}\partial^{\beta}\big(\vec s^{\nu}(\vec k,t)^\dagger \, \partial^{\gamma} \vec s^{\nu}(\vec k,t)\big) \;,
\end{equation}
which involves the eigenvectors $\vec s^{\nu}(\vec k,t)$
and the quasienergies $\epsilon^\nu(\vec k, t)$ of the different bands of the Floquet-Bloch propagator $U(\vec k, t)$.
Both quantities are obtained from diagonalization of $U(\vec k,t)$ as
\begin{equation}
U(\vec k, t) = \sum_{\nu=1}^n e^{-\ii \epsilon^\nu(\vec k, t)} | \vec s^{\nu}(\vec k,t) \rangle\langle \vec s^{\nu}(\vec k,t)| \;.
\end{equation}
For the above expression to make sense, we assume continuous quasienergies $\epsilon^\nu(\vec k, t)$.

In Eq.~\eqref{Berry}, $\epsilon_{\alpha\beta\gamma}$ is the antisymmetric Levi-Civita tensor, the indices $\alpha$, $\beta$, $\gamma$ run over permutations of the parameters $k_1, k_2, t$ of $U(\cdot$), and summation over repeated indices is implied.
 In all expressions, e.g., for $F^\nu_3$, we choose $t$  as the third coordinate.
The integration is over one period $0 \le t \le T$ and over the two-dimensional Brillouin zone $\mathcal B$.
The invariant $W_3(\epsilon)$ depends on the quasienergy $\epsilon$ within a gap through the second term,
the boundary term at $t=T$, where the branch cut of the complex logarithm $\log_{\epsilon}(\cdot)$ is chosen along the line from zero through $e^{-\ii \epsilon}$. 

The above expression, which is the starting point for the construction of the algorithm in Ref.~\onlinecite{HAF17},
is not fully suitable for the present study because it is formulated with respect to an absolute reference point $\epsilon = 0$. Instead, we seek an expression where all quantities are computed relative to the quasienergy $\epsilon$ of the gap under consideration.
 
To obtain this expression, note that the divergence of the Berry curvature $F_{\alpha}^{\nu}(\vec k,t)$ is non-zero only~\cite{Berry45} at a degeneracy point $\vec d_i$ of $U(\vec k,t)$. At such a point, it is $\partial^{\alpha} F_{\alpha}^{\nu}(\vec k_i,t_i)=C^{\nu}(\vec d_i) \delta(\vec k-\vec k_i,t-t_i)$,
where $C^{\nu}(\vec d_i) =\oint_{\mathcal S(\vec d_i)} F_{\alpha}^{\nu} \; \mathrm dS^{\alpha}$ with a small surface around $\vec d_i$.
The quantity $C^{\nu}(\vec d_i)$ is an integer, which can be interpreted as the topological charge of the degeneracy point in band $\nu$ (cf. Ref.~\onlinecite{1367-2630-17-12-125014}).
The net charge of a degeneracy point is zero, that is $C^{\nu}(\vec d_i) = - C^{\mu}(\vec d_i)$ for the two bands $\mu$, $\nu$ that touch at $\vec d_i$.

We can now replace the first term in Eq.~\eqref{W3_old} by a sum over all degeneracy points $i$.
Each degeneracy point gives a contribution of the form
\begin{equation}
 C^{\nu}(\vec d_i) (\epsilon^{\nu}(\vec k_i, t_i) + \Delta_i)  +
 C^{\mu}(\vec d_i) (\epsilon^{\mu}(\vec k_i, t_i)  + \Delta_i) \;,
\end{equation}
where we can include a shift $\Delta_i$ that cancels because of $C^{\nu}(\vec d_i) = - C^{\mu}(\vec d_i)$.
We choose $\Delta_i$ such that $\epsilon^{\nu}(\vec k_i,t_i)+ \Delta_i=\left \lceil{(\epsilon^{\nu}(\vec k_i,t_i)-\epsilon)/(2\pi)} \right \rceil$, with the ceiling function $\lceil \cdot \rceil$ (i.e., rounding up to the next integer). Then, it is also $\epsilon^{\mu}(\vec k_i,t_i)+ \Delta_i=\left \lceil{(\epsilon^{\mu}(\vec k_i,t_i)-\epsilon)/(2\pi)} \right \rceil$ because at a degeneracy point $\epsilon^{\nu}(\vec k_i,t_i)$ and $\epsilon^{\mu}(\vec k_i,t_i)$ differ by a multiple of $2 \pi$.

For the second term in Eq.~\eqref{W3_old}, 
we note that the factor involving the quasienergies does not depend on $\vec k$ when $\epsilon$ is in a gap.
Therefore, it can be pulled out of the integral.
Evaluation of the complex logarithm, with the branch cut at the right position, gives
\begin{equation}\label{branch}
\frac{\ii \log_{\epsilon} e^{-\ii \epsilon^{\nu}(\vec k,T)}- \epsilon^{\nu}(\vec k,T)}{2\pi} =
 \left \lceil{\frac{\epsilon-\epsilon^{\nu}(\vec k,T)}{2\pi}}\right \rceil \;.
\end{equation}
For the third term in Eq.~\eqref{W3_old}, we have similarly that $\epsilon^{\nu}(\vec k,0)$ does not depend on $\vec k$ and, because of $U(\vec k,0) = 1$, is in fact a multiple of $2 \pi$.

Now we can sum the contribution of all degeneracy points to one band $\nu$, and find
\begin{equation}
 C^\nu - C^\nu_0 =  \iint\limits_{\mathcal B}  F_3^{\nu}( \vec k, T ) - F_3^{\nu}( \vec k, 0 ) \, \mathrm dk_1 \mathrm dk_2  = \sum_{i=1}^\mathrm{dp} C^{\nu}(\vec d_i) \;,
\end{equation}
where \mbox{$C^\nu = \iint\limits_{\mathcal B}  F_3^{\nu}( \vec k, T ) \, \mathrm dk_1 \mathrm dk_2 $} and \mbox{$C^\nu_0 = \iint\limits_{\mathcal B}  F_3^{\nu}( \vec k, 0) \, \mathrm dk_1 \mathrm dk_2$} are the Chern numbers of band $\nu$ at the final time $t=T$ and initial time $t=0$.

Putting everything together, we arrive at
\begin{multline}\label{W3_new}
  W_3(\epsilon) = \sum_{\nu=1}^n \Bigg[\sum_{i=1}^{\text{dp}} N^{\nu}(\epsilon,\vec d_i) C^{\nu}(\vec d_i) \\ + \left \lceil{\frac{\epsilon-\epsilon^{\nu}(\vec k,T) + \epsilon^\nu(\vec k, 0)}{2\pi} }\right \rceil C_0^{\nu}\Bigg]\; ,
\end{multline}
with 
\begin{equation}
N^\nu(\epsilon, \vec d_i) = \left\lceil \frac{\epsilon^\nu(\vec k_i, t_i) - \epsilon}{2 \pi} \right\rceil + \left\lceil \frac{\epsilon-\epsilon^\nu(\vec k, T)}{2 \pi} \right\rceil \;.
\end{equation}
Note that these expressions are invariant under shifts $\epsilon^\nu(\cdot) \mapsto  \epsilon^\nu(\cdot) + 2 \pi m$ of the quasienergies of a band by multiples of $2 \pi$, as it should.
We can especially choose $\epsilon^\nu(\vec k,0) = 0$, if we prefer, for example as in Fig.~1 in the main text.
For the sake of brevity, we also drop the last term
and set $C_0^{\nu}=0$ in the main text, as if all bands were topologically trivial at $t=0$.

One might want to note the similarity of Eq.~\eqref{W3_new} to Eq.~(4.4) in Ref.~\onlinecite{HAF17},
which is the basis of the algorithm presented there.
Owing to this similarity, evaluation of the above expression, and also of the $\mathbb Z_2$-invariants defined in the main text,
is possible with that algorithm.

Let us finally remark that the expression for the $W_3$-invariant given in Eq.~(9) of Ref.~\onlinecite{1367-2630-17-12-125014}
can be recovered from our Eq.~\eqref{W3_new} if we adopt the same ordering of the Floquet bands in a ``natural quasienergy Brillouin zone''.
Specifically, we have to (a) set $\epsilon^{\nu}(\vec k,0)=0$, (b) impose the ordering condition: $\epsilon^{\nu}(\vec k,t)\le \epsilon^{\nu'}(\vec k,t)$ for $\nu < \nu'$,
and (c) assume that $\epsilon^n(\vec k,t) - \epsilon^1(\vec k,t)  \le 2 \pi$.

Now suppose that the gap at $\epsilon$ separates Floquet bands $m$, $m+1$, that is $\epsilon^{m}(\vec k,T) < \epsilon < \epsilon^{m+1}(\vec k,T)$.
In this case, Eq.~\eqref{W3_new} reduces to
\begin{equation}\label{AppB1}
W_3(\epsilon)=\sum_{\nu=1}^m C^{\nu}+\sum_{\nu=1}^n \sum_{i=1}^{\text{dp}}\left\lceil \frac{\epsilon^\nu(\vec k_i, t_i) - \epsilon}{2 \pi} \right\rceil C^{\nu}(\vec d_i) \; .
\end{equation}
In this expression, the contributions from a degeneracy point $\vec d_i$ that occurs between two bands $1 \le \mu < \mu+1 \le n$, that is for $\epsilon^\mu(\vec k_i, t_i) = \epsilon^{\mu+1}(\vec k_i, t_i)$, cancel:
The ceiling function $\lceil \cdot \rceil$ has the same value for $\nu \in \{ \mu, \mu+1\}$, but $C^\mu(\vec d_i) = - C^{\mu+1}(\vec d_i)$.
Only the degeneracy points that occur between bands $1$, $n$, which fulfill $\epsilon^1(\vec k_i, t_i) = \epsilon^n(\vec k_i, t_i) - 2\pi $, contribute: Now $\lceil \cdot \rceil = 0$ for $\nu = 1$, but $\lceil \cdot \rceil = 1$ for $\nu = n$.
In Ref.~\onlinecite{1367-2630-17-12-125014}, these degeneracy points are called ``zone-edge singularities''.
We thus obtain, under the above assumptions, an expression of the form
\begin{equation}\label{AppB2}
W_3(\epsilon)=\sum_{\nu=1}^m C^{\nu}+\sum_{i=1}^{\mathrm{dp}} C^{n}(\vec d_i) \; ,
\end{equation}
which is, up to notational differences, Eq.~(9) from Ref.~\onlinecite{1367-2630-17-12-125014}.
We can thus recognize this equation as a special case of the more general Eq.~\eqref{W3_new}.

\section{Chiral symmetry with a momentum shift $\vec k_\pi$}
\label{app:Ch}

\begin{figure*}
\hspace*{\fill}
\raisebox{20pt}{\includegraphics[scale=1]{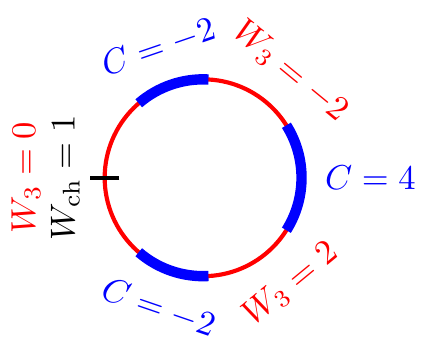}}
\hspace*{\fill}
\includegraphics[scale=0.3]{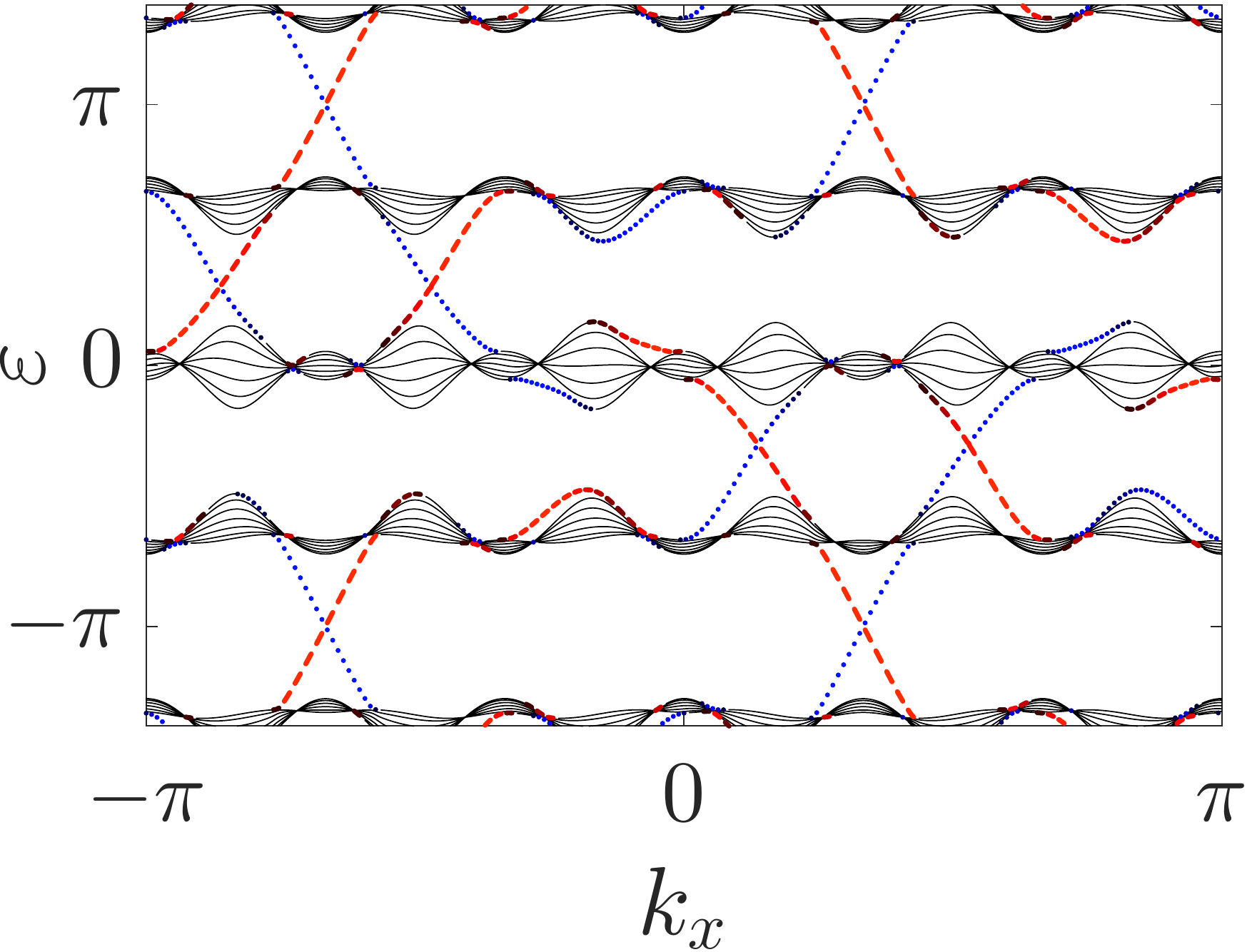}
\hspace*{\fill}
\includegraphics[scale=0.3]{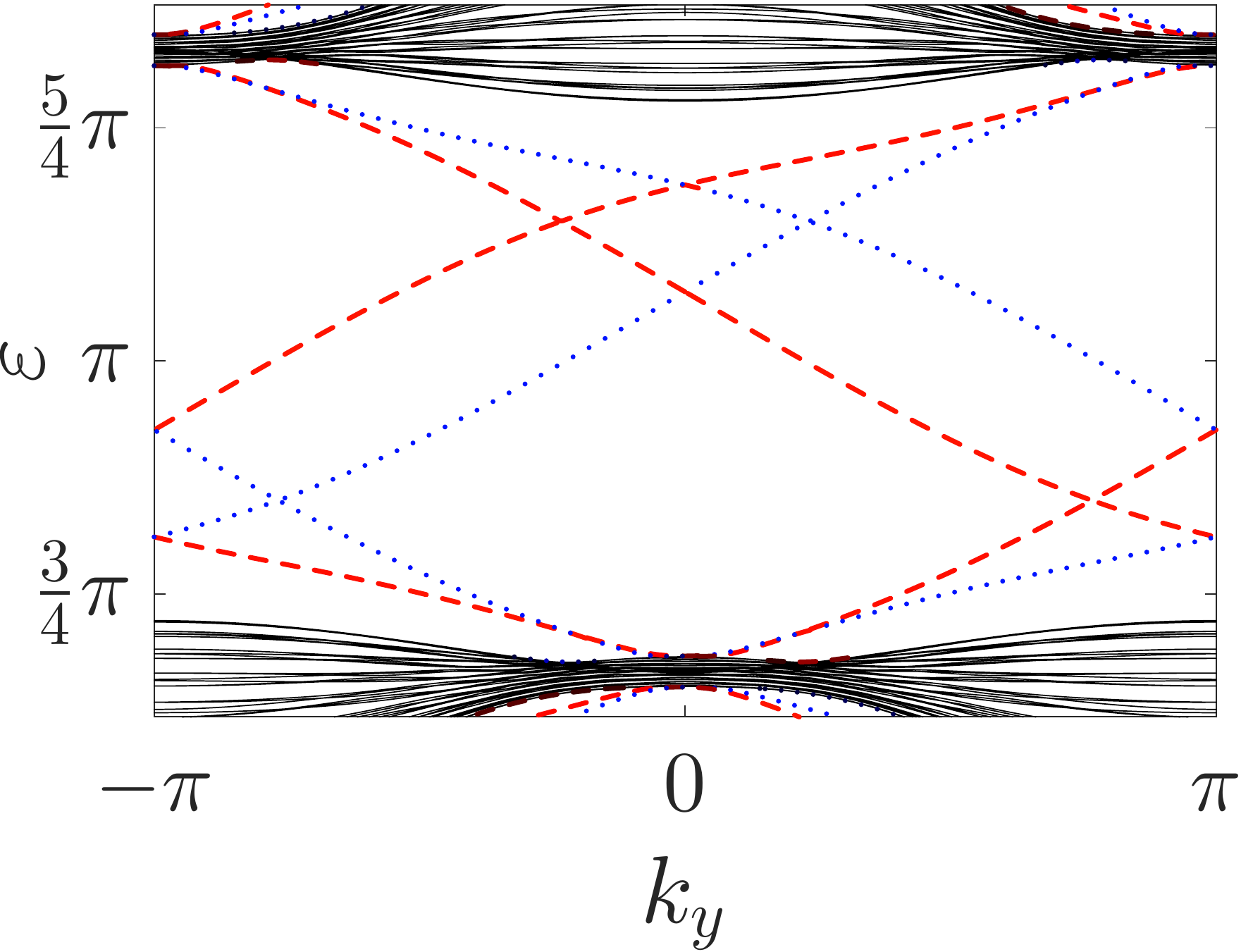}
\hspace*{\fill}
\caption{Same as Fig.~\ref{fig:CH}, now for the periodically kicked version of the Harper model~\eqref{CH_Ham} as in Ref.~\onlinecite{PhysRevB.93.075405} with $\tilde J_x = \pi$. 
Left panel: Included are the Chern numbers of each band,
and the $W_3$ and $W_\mathrm{ch}$-invariants in each gap.
Central and right panel:  Bands and boundary states on a semi-infinite strip along the $x$ and $y$-axis.
In the right panel, we show only the gap at $\epsilon = \pi$ for better visibility. For both boundary configurations, one pair of symmetry-protected topological boundary states exists in the gap at $\epsilon=\pi$ in accordance with $W_{\mathrm{ch}}(\pi) \ne 0$ in the left panel.}
\label{Fig:appCh1}
\end{figure*}

In Eq.~\eqref{CH_Symm} chiral symmetry is defined with a $\vec k \mapsto \vec k + \vec k_\pi$ momentum shift,
which differs from the standard definition in the literature~\cite{PhysRevB.93.115429, PhysRevB.96.155118,PhysRevB.96.195303},
\begin{equation}\label{CH_Symm_Lit}
 \tilde H_\mathrm{ch}(\vec k, t) = - S \tilde H_\mathrm{ch}(\vec k ,T-t) S^{-1} \;,
\end{equation}
that does not involve a momentum shift.

The origin of the momentum shift in Eq.~\eqref{CH_Symm} is a bipartite even-odd sublattice symmetry assumed there. Specifically, we consider the original lattice, whose units cells are enumerated by two indices $(i,j)$, as being composed of the sublattices of even ($i+j \equiv 0 \mod 2$) and odd ($i+j \equiv 1 \mod 2$) unit cells.
If the chiral symmetry operator includes an alternating sign flip for every second unit cell of the lattice, 
say for the odd unit cells,
the sign flip translates into the shift $\vec k \mapsto \vec k + \vec k_\pi$ for the Bloch Hamiltonian.

We can now consider the Bloch Hamiltonian for a $2 \times 2$ unit cell
that comprises four unit cells of the original lattice. If we enumerate these four unit cells in the obvious way,
say in the order $(2i, 2j)$, $(2i +1, 2j)$, $(2i, 2j+1)$, $(2i+1, 2j+1)$,
the new Bloch Hamiltonian has the $4 \times 4$ block form
\begin{equation}\label{HDouble}
 \hat H (\vec k,t) = \begin{pmatrix} H_\mathrm{loc} & H_x & H_y & H_d \\ H_x &   H_\mathrm{loc} & H_d & H_y \\ H_y & H_d & H_\mathrm{loc}  & H_x \\ H_d & H_y & H_x & H_\mathrm{loc} \end{pmatrix} \;.
\end{equation}
It contains diagonal blocks $H_\mathrm{loc} \equiv H_\mathrm{loc}(t)$ for terms within a unit cell, and the off-diagonal blocks $H_{x/y} \equiv H_{x/y}(\vec k,t)$ for hopping along the two lattice axes and $H_d \equiv H_d(\vec k,t)$ for diagonal hopping.
For a Hamiltonian with only nearest-neighbor hopping, the blocks $ H_d \equiv 0$ vanish.
The equality of the diagonal and off-diagonal blocks incorporated into Eq.~\eqref{HDouble} follows from the translational symmetry of the original Hamiltonian.
Note that we do not assume additional geometric symmetries, and allow for $H_x \ne H_y$.

\begin{figure*}
\hspace*{\fill}
\raisebox{20pt}{\includegraphics[scale=1]{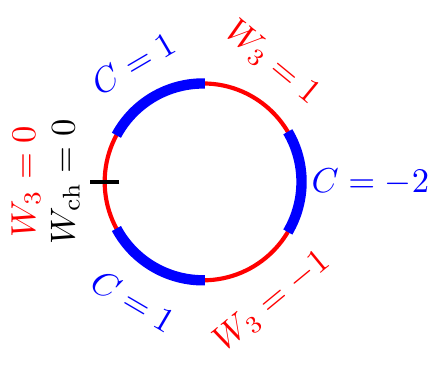}}
\hspace*{\fill}
\includegraphics[scale=0.3]{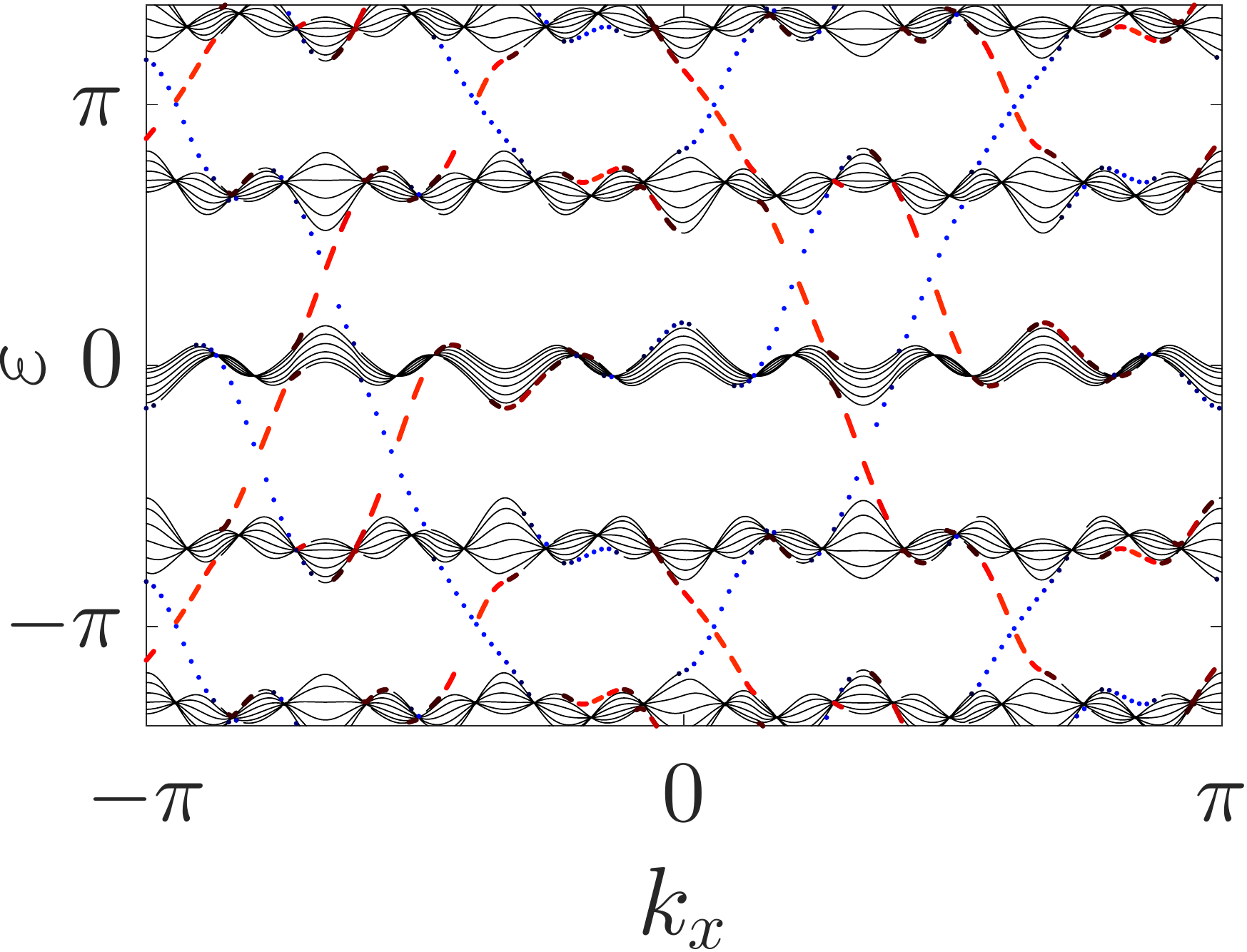}
\hspace*{\fill}
\includegraphics[scale=0.3]{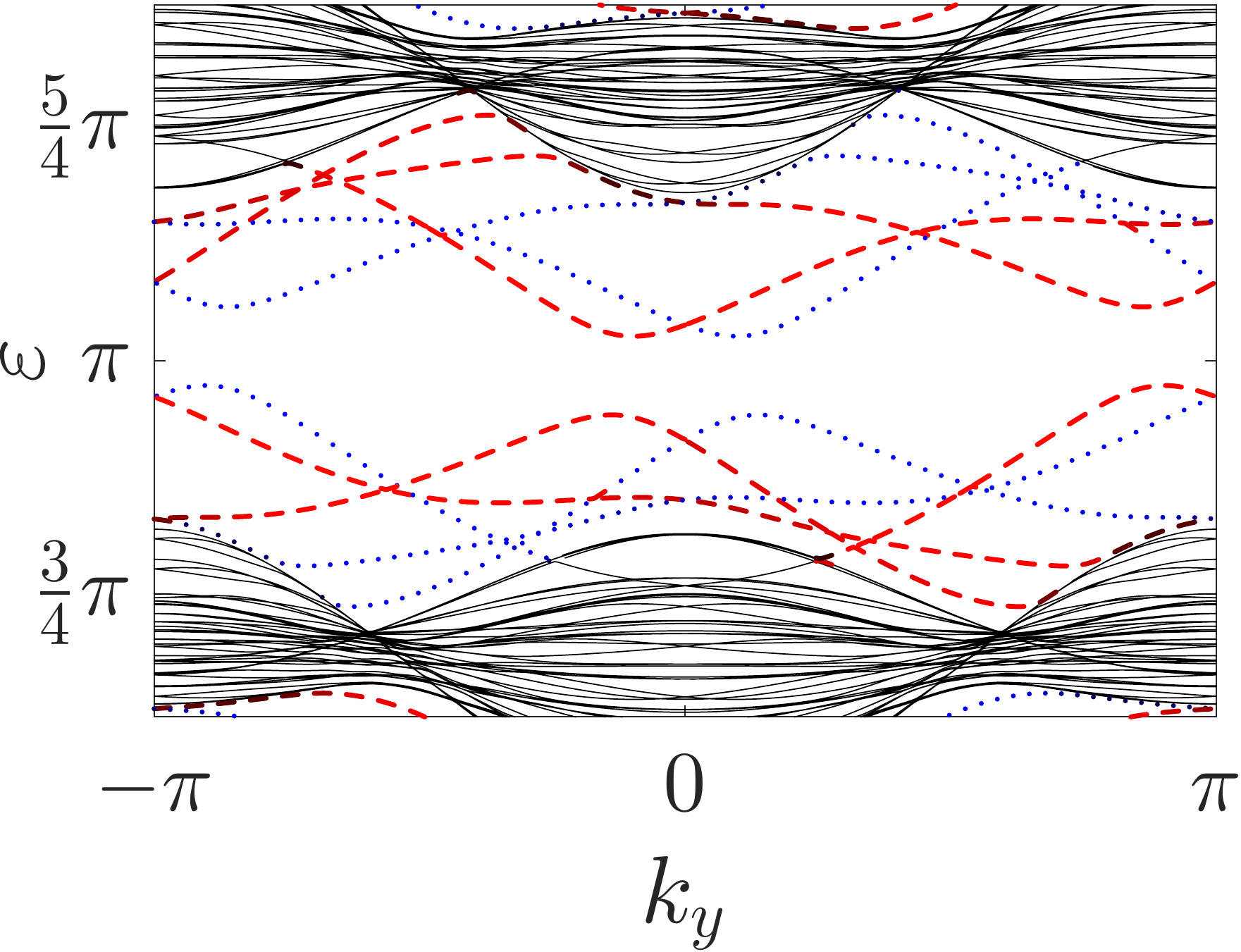}
\hspace*{\fill}
\caption{Same as Fig.~\ref{Fig:appCh1}, now for $\tilde J_x=3/2\pi$. 
Central panel: On a boundary along the $x$-axis, two pairs of boundary states with opposite chirality exist in the gap at $\epsilon=\pi$. The two pairs are not symmetry-protected and can annihilate each other.
Right panel: On a boundary along the $y$-axis, no boundary state crossing the gap at  $\epsilon=\pi$ exists.
For both boundary configurations, the number of boundary states (taken modulo $2$) agrees with the value $W_{\mathrm{ch}}(\pi)=0$ in the left panel.}
\label{Fig:appCh2}
\end{figure*}

The chiral symmetry operator for $\hat H(\vec k, t)$,
\begin{equation}
 \hat S = \begin{pmatrix} S &  \\  & -S \\ & & -S & \\ & & & S \end{pmatrix} \;,
\end{equation}
 is block-diagonal. The plus and minus signs of the entries correspond to the alternating sign flip on the even-odd sublattice structure.
If the original Hamiltonian $H(\vec k, t)$ satisfies Eq.~\eqref{CH_Symm},
we have $\hat S \hat H(\vec k,t) \hat S^{-1} = - \hat H(\vec k,T-t)$ for the new Hamiltonian.
Therefore, $\hat H(\vec k,t)$ fulfills the standard chiral symmetry relation~\eqref{CH_Symm_Lit}.

To obtain $\hat H(\vec k,t)$ we have considered only translations by an even number of sites on the original lattice. $\hat H(\vec k,t)$ inherits additional symmetries from translations by an odd number of sites.
The two symmetry operators are 
\begin{equation}
 \hat T_x = \begin{pmatrix} 0 & \mathbbm 1 & 0 & 0 \\  \mathbbm 1 & 0 &0 &0 \\ 0 & 0 & 0 & \mathbbm 1 \\ 0 & 0 & \mathbbm 1 & 0 \end{pmatrix} \;,
 \quad
\hat T_y = \begin{pmatrix} 0 & 0 & \mathbbm 1 & 0 \\  0 & 0 & 0 & \mathbbm 1 \\ \mathbbm 1 & 0 & 0 & 0 \\ 0 & \mathbbm 1 & 0 & 0 \end{pmatrix} \;.
\end{equation}
Since $[\hat T_x, \hat T_y]=0$, only one of the two symmetry operators is needed below.
Note that if we want to interpret $\hat T_{x/y}$ as a translation on the original lattice some
prefactors $\sim e^{\ii k_i}$ must be included, but since the prefactors cancel trivially in all relations we have dropped them here.
We have $[\hat T_{x/y}, \hat H (\vec k,t)] = 0$,
but $\hat S \, \hat T_{x/y} \, \hat S^{-1} = - \hat T_{x/y}$.

The above relations carry over to the Floquet-Bloch propagator $\hat U(\vec  k, t$) associated to $\hat H(\vec k,t)$.
We have $S \hat U_{\star}(\vec k,t)S^{-1} = \hat U_{\star}^\dagger(\vec k, t)$,
and $[\hat T_{x/y}, \hat U_\star (\vec k,t)] = 0$.

Now let us assume that $|\psi\rangle$ is an eigenstate of $\hat U_\star(\vec k,t)$, to the quasienergy $\epsilon$.
The state $| \zeta \rangle = \hat S |\psi\rangle$ is an eigenstate of $\hat U_\star(\vec k,t)$ to the negative quasienergy $-\epsilon$.
Now if $\epsilon = 0,\pi$ the states $|\psi\rangle$, $|\zeta\rangle$ are degenerate.
For the original Hamiltonian, with symmetry relation~\eqref{CH_Symm}, degenerate quasienergies occur at momenta $\vec k$, $\vec k + \vec k_\pi$.
For the Hamiltonian $\hat H$, with symmetry relation~\eqref{CH_Symm_Lit}, degeneracies occur at the same momentum $\vec k$.

For time-reversal symmetry, where a similar situation occurs at the IM, Kramers' theorem implies the orthogonality of the two degenerate states, and thus the symmetry-protection of the corresponding topological phases.
For chiral symmetry, the eigenstates can be classified by means of the symmetry operator $\hat T_x$ (or $\hat T_y$), as $\hat T_x|\psi\rangle = \pm |\psi\rangle$.
Now $|\zeta\rangle$ is also an eigenstate of $\hat T_x$, with the negative eigenvalue $\hat T_x|\zeta\rangle = \mp |\zeta\rangle$.
This observation implies the orthogonality of $|\psi\rangle$ and $|\zeta\rangle$.

Therefore, the situation for chiral symmetry is, although for different reasons, analogous to the situation for time-reversal symmetry:
In both cases symmetry-protected topological phases exist because degenerate states occur only in orthogonal pairs.
We repeat that without the momentum shift $\vec k_\pi$ no such argument is possible, and we should not expect that a symmetry-protected topological phase exists in $2+1$ dimensional systems with that type of chiral symmetry.

To support these findings with additional numerical evidence
we show in Figs.~\ref{Fig:appCh1},~\ref{Fig:appCh2} 
invariants and boundary states of the periodically kicked Harper model introduced in Ref.~\onlinecite{PhysRevB.93.075405} for the study of topological phases with chiral symmetry.
This model is equal to the Harper model of Eq.~\eqref{CH_Ham}, now with the time-dependence 
$J_x(t)=\tilde J_{x}\sum_{m=-\infty}^{\infty} \delta(t-mT/2)$.

In Fig.~\ref{Fig:appCh1}, with parameters $\alpha=1/3$, $\tilde J_x=\pi$, $J_y=\pi/3$ that correspond to the central panel of Fig.~1 in Ref.~\onlinecite{PhysRevB.93.075405}, we observe one pair of boundary states with opposite chirality in accordance with the non-zero value $W_\mathrm{ch}(\pi)$ of the $W_\mathrm{ch}$-invariant.
This pair exists independently of the boundary orientation.
Note that in the gaps between $\epsilon = 0, \pi$, which have no special significance for chiral symmetry,
the number of unpaired boundary states is given by the $W_3$-invariant.

Having changed the parameter $\tilde J_x$ to $\tilde J_x=3/2\pi$ in Fig.~\ref{Fig:appCh2}, which corresponds to the right panel of Fig.~1 in Ref.~\onlinecite{PhysRevB.93.075405}, gaps have closed and reopened.
The values of the invariants have changed, and now $W_\mathrm{ch}(\pi)=0$.
Since $W_\mathrm{ch}(\pi)$ is a $\mathbb Z_2$-invariant we expect an even number of pairs of boundary states with opposite chirality.
Indeed, we observe two pairs on a boundary along the $x$-axis (central panel), and zero pairs on a boundary along the $y$-axis (right panel). The two pairs are not protected, and could be annihilated by variation of additional model parameters~\cite{PhysRevB.93.075405}.

These results agree with Ref.~\onlinecite{PhysRevB.93.075405}, and with our statements in the main text.
In particular, we observe the existence or absence of symmetry-protected boundary states in dependence on the value of the $\mathbb Z_2$-invariant $W_\mathrm{ch}(\epsilon)$, but independently of the boundary orientation.

\section{Degeneracy points and the Kane-Mele invariant}
\label{app:KM}

In the time-reversal symmetric case we can define an effective Brillouin zone $\mathcal E$ such that either $\vec k \in \mathcal E$ or $- \vec k \in \mathcal E$.
Then, the sum over half of the degeneracy points in Eq.~(6) for the $W_\mathrm{tr}$-invariant in the main text can be performed by counting exactly the degeneracy points $\vec d_i$ with $\vec k_i \in \mathcal E$.
Including the time coordinate, these degeneracy points lie in the 
box $\mathcal {BX} = \mathcal E \times [0,T]$.

Now consider a single Kramers pair of bands $2\nu-1$, $2\nu$. The sum over all degeneracy points of this pair can be written as
\begin{equation}
\begin{split}
& \sum_{i=1}^{\text{dp}/2} C^{2\nu-1}(\vec d_i)+ C^{2\nu}(\vec d_i) = \\
& \iiint\limits_{\mathcal{BX}} \partial^{\alpha} F_{\alpha}^{2\nu-1}(\vec k,t) + \partial^{\alpha} F_{\alpha}^{2\nu}(\vec k,t) \; \mathrm dk_1 \mathrm dk_2 \mathrm dt \;.
\end{split}
\end{equation}
With Gauss's theorem we can convert this integral into an integral over the surface of the box $\mathcal{BX}$,
which is the union of the two faces $\mathcal F_0 = \mathcal E \times \{0 \}$, $\mathcal F_T = \mathcal E \times \{T \}$ and the cylinder $\mathcal C = \partial \mathcal E \times [0,T]$ that contains the points on the boundary curve $\partial \mathcal E$ of $\mathcal E$.

With Stokes' theorem, the integral of the Berry curvature $F_\alpha^{2 \nu -1}$, $F_\alpha^{2 \nu}$  over $\mathcal C$ can be converted further into a line integral of the Berry connection $A^{\alpha, 2\nu-1}$, $A^{\alpha, 2\nu-1}$ over the two curves $\partial\mathcal E \times \{0\}$, $\partial\mathcal E \times \{T\}$. 
Recall that in terms of the eigenvectors of $U_\star(\cdot)$, it is
$A^{\alpha,\nu}(\vec k,t)=\frac{1}{2\pi \ii}\big[\vec s^{\nu}(\vec k,t)\big]^{\dagger} \partial^{\alpha} \vec s^{\nu}(\vec k,t)$.
Since the Berry connection is gauge-dependent, we here have to impose a time-reversal constraint
on $\partial \mathcal E$, namely
\begin{equation}
\begin{split}
\vec s^{2\nu-1}(-\vec k,t)&=\Theta \, \vec s^{2\nu}(\vec k, t) \; ,\\
\vec s^{2\nu}(- \vec k,t)&=-\Theta \, \vec s^{2\nu-1}(\vec k,t) \;,
\end{split}
\label{TR_gauge}
\end{equation}
to obtain the Kane-Mele invariants in a manner analogously to Ref.~\onlinecite{PhysRevB.74.195312}.
Note that the $\vec s(\cdot)$ in this expression are the eigenvectors of the time-symmetrized propagator $U_\star(\cdot)$,
such that the time argument $t$ is unchanged while $\vec k$ is flipped.

We arrive at the relation (everything taken modulo two)
\begin{equation}
\begin{split}
& \sum_{i=1}^{\text{dp}/2} C^{2\nu-1}(\vec d_i)+ C^{2\nu}(\vec d_i) \\
& \equiv \iint\limits_{\mathcal E} F_{\alpha}^{2\nu-1}(\vec k,T) + F_{\alpha}^{2\nu}(\vec k,T) \; \mathrm dk_1 \mathrm dk_2 \\
& \qquad - \int_{\partial \mathcal E} A^{\alpha,2\nu-1}(\vec k,T) + A^{\alpha,2\nu}(\vec k,T)\; \mathrm dk_{\alpha}  \\
& \qquad - \iint\limits_{\mathcal E} F_{\alpha}^{2\nu-1}(\vec k,0) + F_{\alpha}^{2\nu}(\vec k,0) \; \mathrm dk_1 \mathrm dk_2 \\
& \qquad + \int_{\partial \mathcal E} A^{\alpha,2\nu-1}(\vec k,0) + A^{\alpha,2\nu}(\vec k,0) \; \mathrm dk_{\alpha}   \\
& \equiv \mathrm{KM}^\nu(T) - \mathrm{KM}^\nu(0)  \;,
\end{split}
\end{equation}
and recognize~\cite{PhysRevB.74.195312}
 the Kane-Mele invariants $\mathrm{KM}^\nu(0)$ and $\mathrm{KM}^\nu(T)$ of the Kramers pair $2\nu-1$, $2\nu$ at $t=0$ and $t=T$.
Therefore, the Kane-Mele invariants can be expressed as the sum over half of the degeneracy points of each Kramers pair.
This observation justifies the corresponding statements in the main text.
For the sake of brevity of the presentation, we there assume $\mathrm{KM}^\nu(0)=0$, as if the Kramers pairs were topologically trivial at $t=0$.

\section{Particle-hole symmetric boundary states on a hexagonal lattice}
\label{app:PH}

\begin{figure}
\hspace*{\fill}
\includegraphics[width=0.9\linewidth]{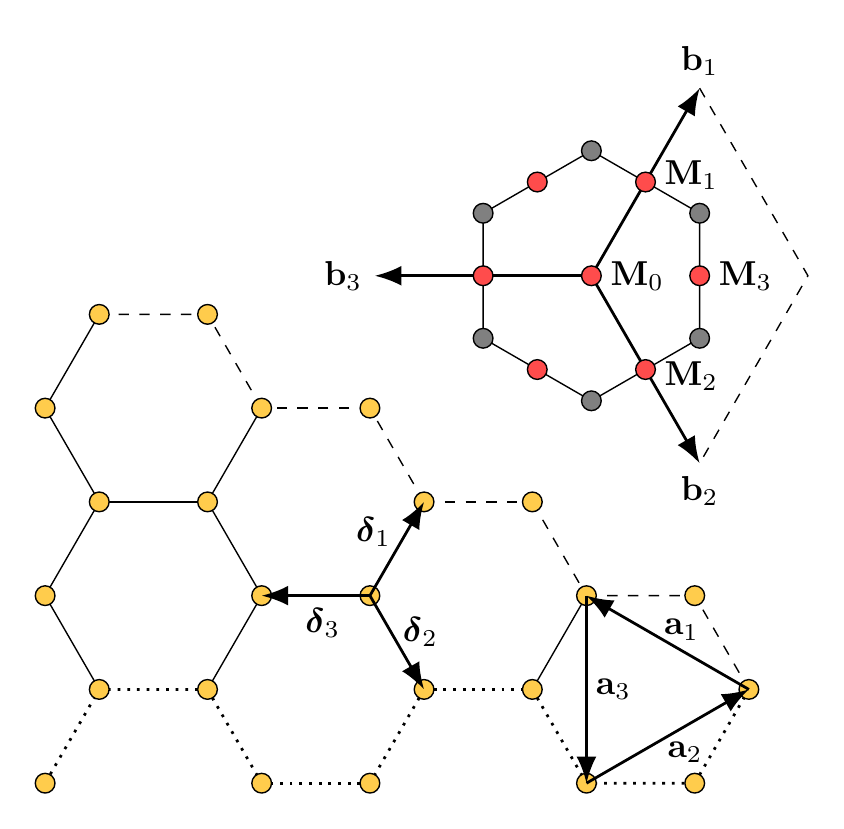}
\hspace*{\fill}
\caption{Primitive lattice vectors $\vec a_i$ and nearest-neighbor vectors $\boldsymbol \delta_i$ on a hexagonal lattice.
Primitive lattice vectors point along the direction of zigzag boundaries (dashed, along $\vec a_1$),
nearest-neighbor vectors along the direction of armchair boundaries (dotted, along $\boldsymbol \delta_3$).
Also shown are the primitive reciprocal lattice vectors $\vec b_i$,
together with red circles that indicate the IM $\vec M_0, \dots \vec M_3$ in the Brillouin zone.
}
\label{fig:Hex}
\end{figure}

For a hexagonal lattice as in Fig.~\ref{fig:Hex},
zigzag boundaries occur along directions given by primitive lattice vectors $\vec a_1$, $\vec a_2$, $\vec a_3$.
Armchair boundaries occur along directions given by nearest-neighbor vectors $\boldsymbol \delta_1$, $\boldsymbol \delta_2$, $\boldsymbol \delta_3$. Note that the primitive translation vector for an armchair boundary is $3\hspace{1pt}\boldsymbol\delta_i$.
In contrast to, say, the situation for a square lattice, both boundary types exist in three inequivalent orientations. This necessitates the more detailed analysis provided here.

To evaluate Eq.~\eqref{W_PH} for each boundary,
we first need to project the IM $\vec M_1, \vec M_2, \vec M_3$ onto the boundary direction (the projection of $\vec M_0$ results in zero).
For zigzag boundaries, we have $\vec a_i \cdot \vec M_i = 0$, and $\vec a_i \cdot \vec M_m = \pi$ for the remaining two IM with $m \ne i$.
For armchair boundaries we obtain essentially the same result:
$3\hspace{1pt} \boldsymbol \delta _i \cdot \vec M_i = 0$, and $3\hspace{1pt} \boldsymbol \delta _i \cdot \vec M_m = \pi$ for $m \ne i$.
Note that all values are given modulo $2 \pi$.
We recognize that for each boundary orientation two IM will contribute in Eq.~\eqref{W_PH} for given momentum $k_{\vec a} = 0, \pi$.

\begin{figure*}
\hspace*{\fill}
\includegraphics[scale=0.28]{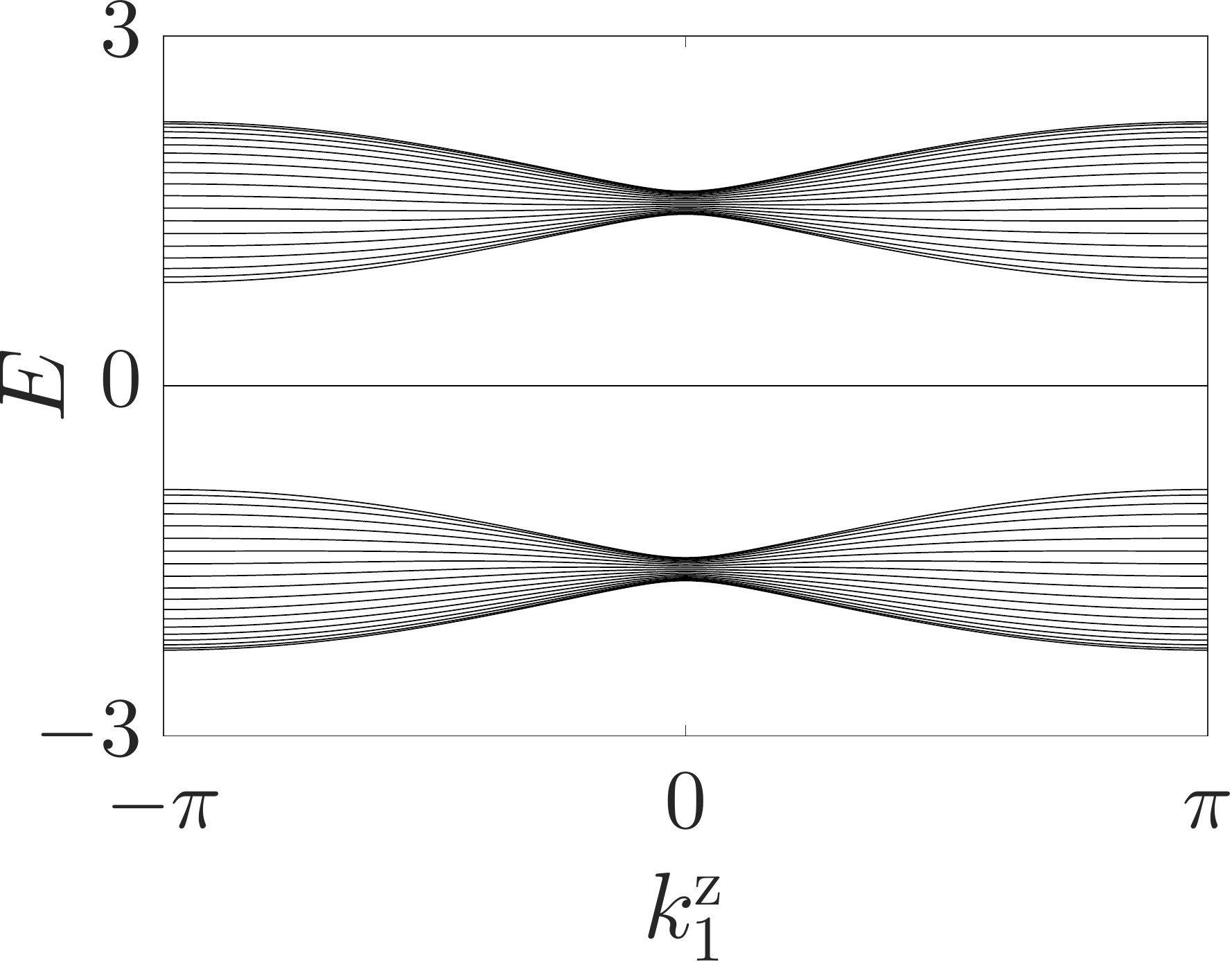}
\hspace*{\fill}
\includegraphics[scale=0.28]{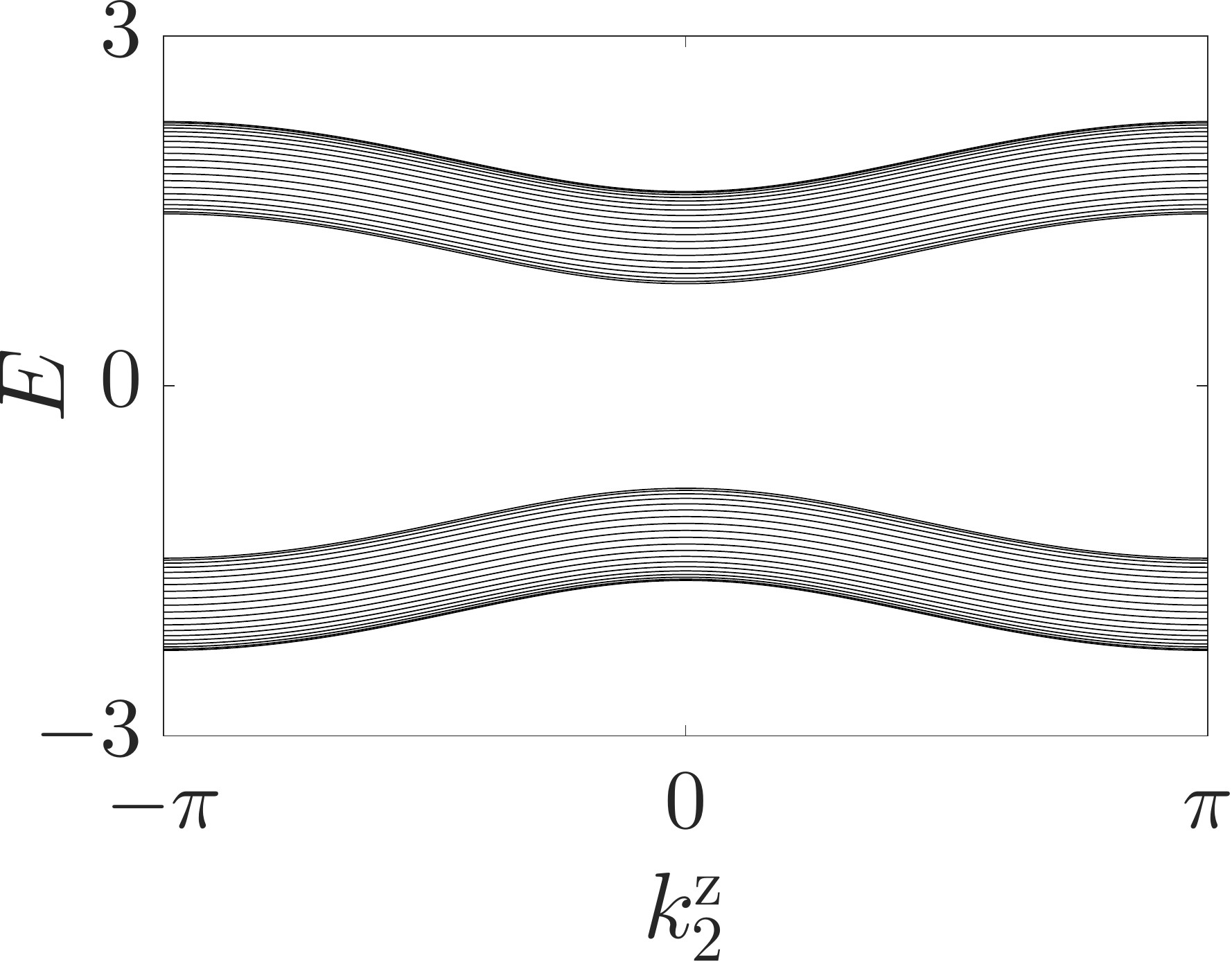}
\hspace*{\fill}
\includegraphics[scale=0.28]{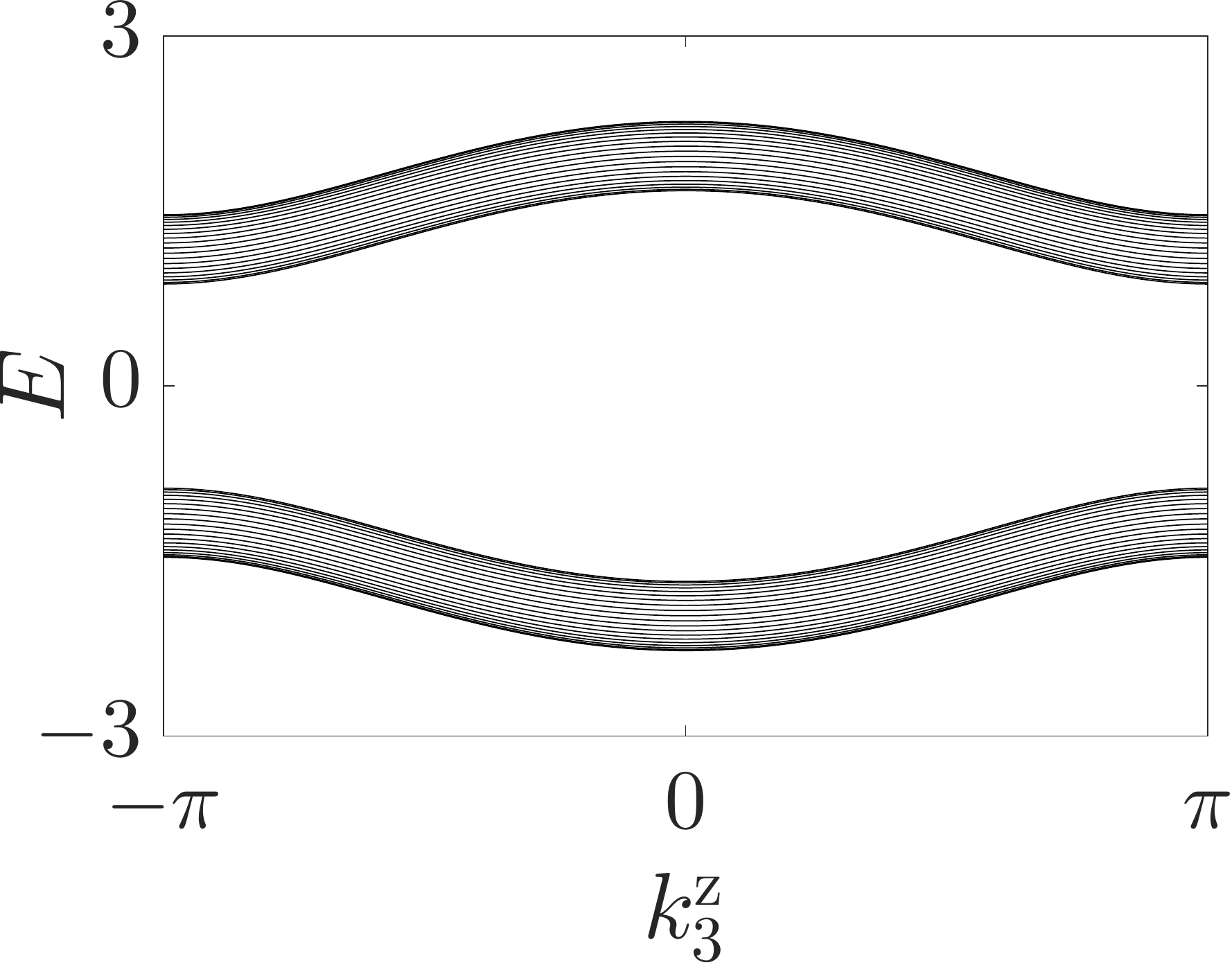}
\hspace*{\fill} \\[1ex]
\hspace*{\fill}
\includegraphics[scale=0.28]{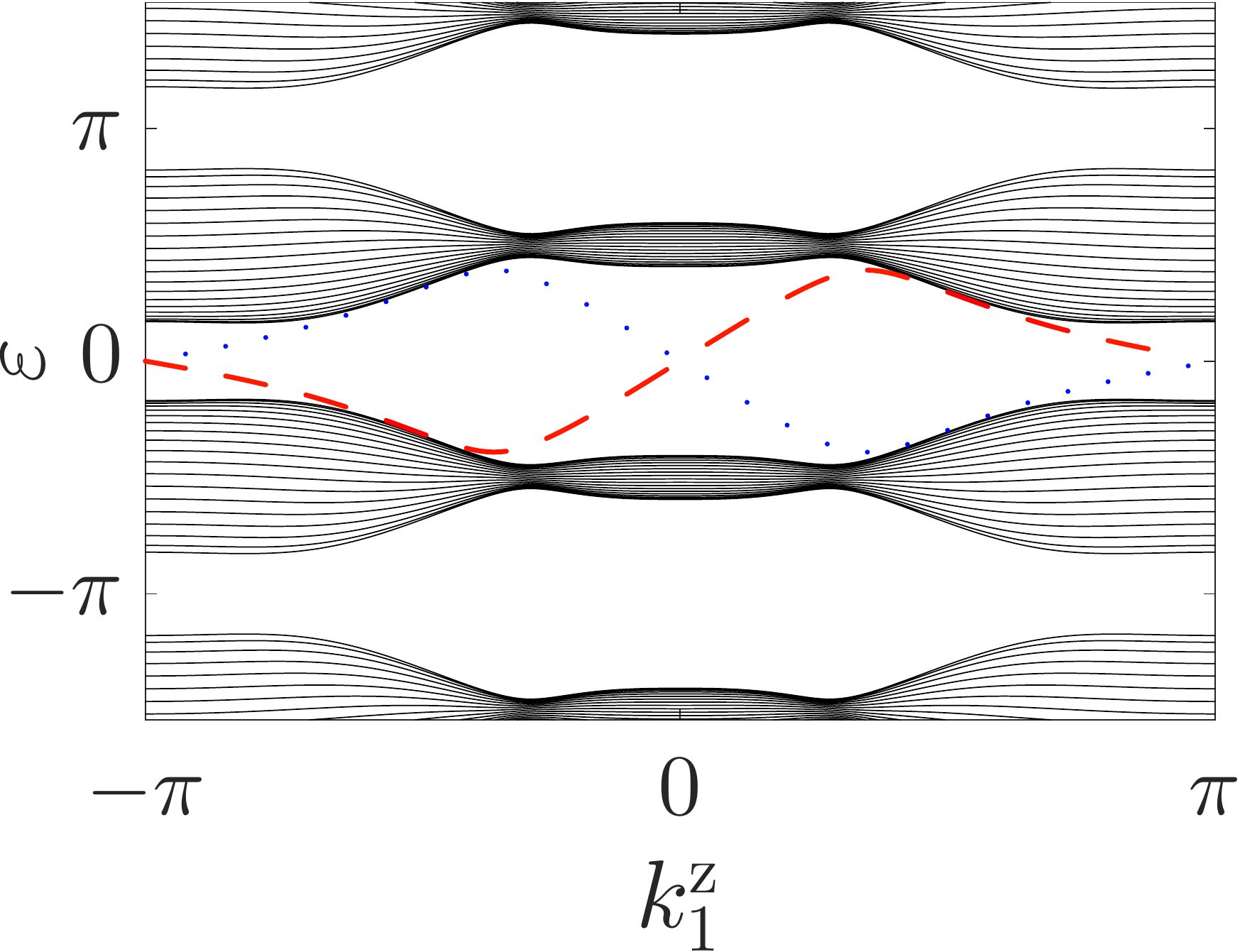}
\hspace*{\fill}
\includegraphics[scale=0.28]{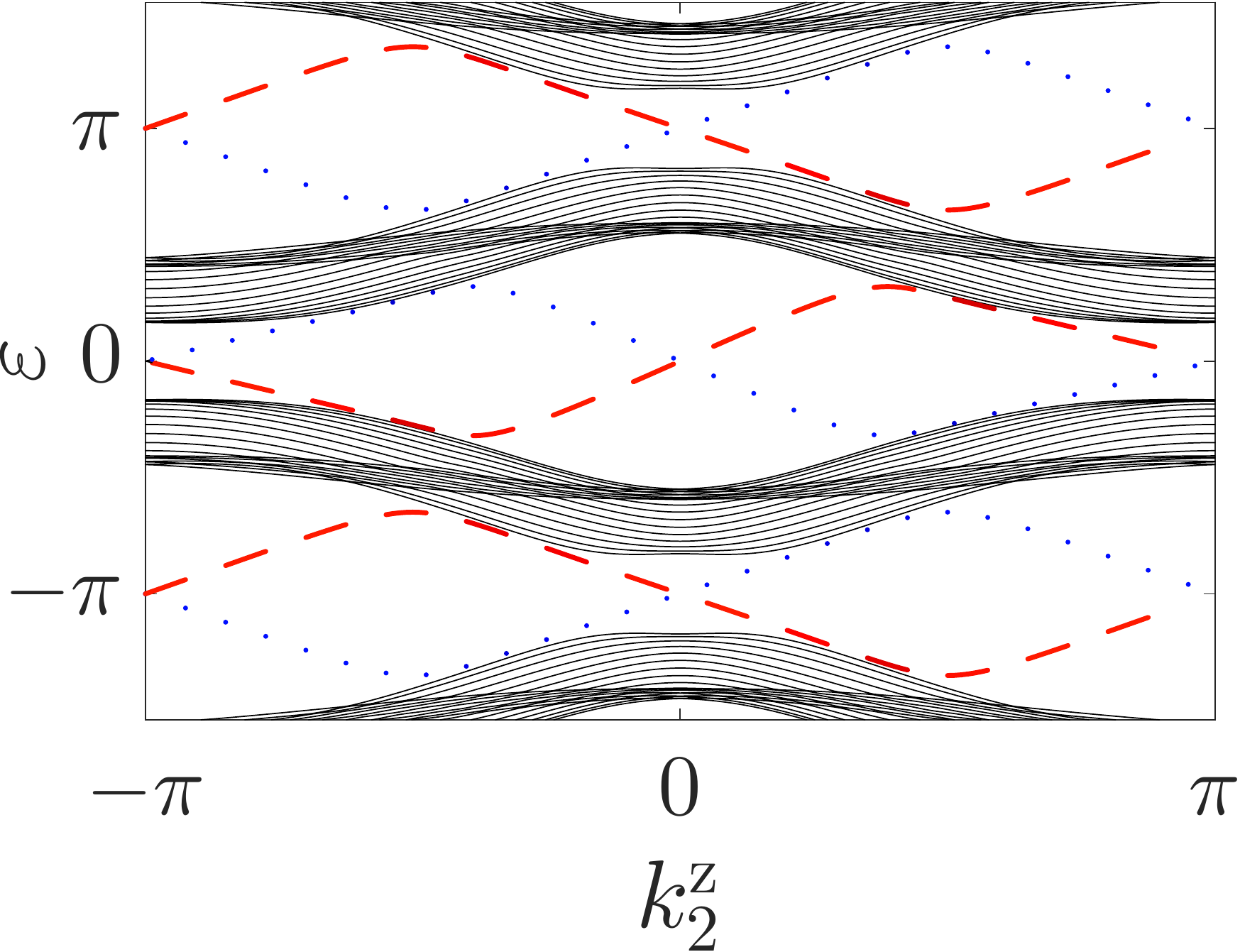}
\hspace*{\fill}
\includegraphics[scale=0.28]{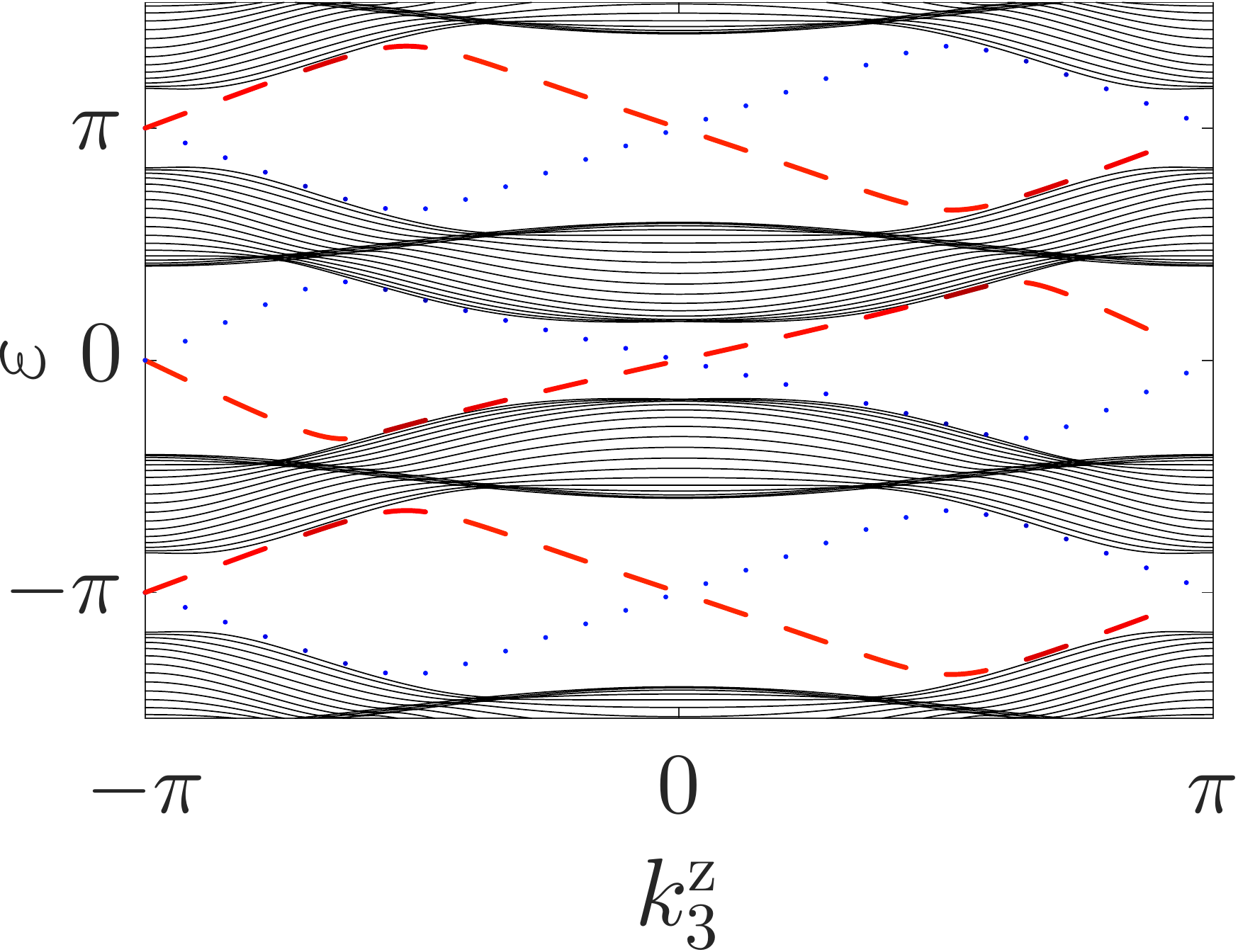}
\hspace*{\fill}
\caption{Initial boundary states of $H_\mathrm{ph}(t=0)$ (top row) and Floquet-Bloch boundaries states at $t=T$ (bottom row) for the 
particle-hole symmetric model~\eqref{PH_Ham}, on zigzag boundaries along the lattice vectors $\vec a_1$, $\vec a_2$, $\vec a_3$ (from left to right). Shown are the energies $E(k^z_i)$ or the quasienergies $\epsilon(k^z_i)$ as a function of the respective momentum $k^z_1$, $k^z_2$, $k^z_3$.
}
\label{fig:zigzag}
\end{figure*}

\begin{figure*}
\hspace*{\fill}
\includegraphics[scale=0.28]{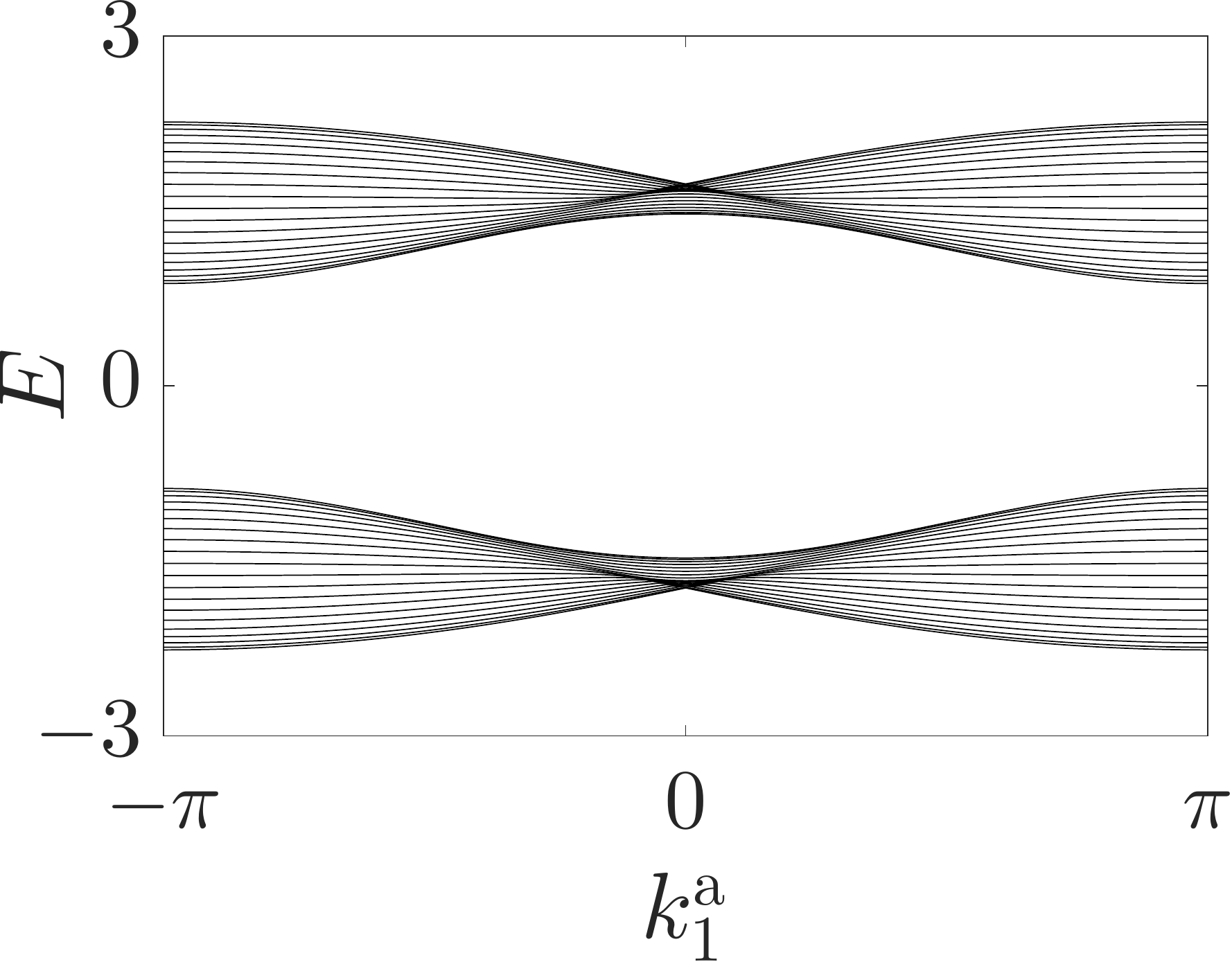}
\hspace*{\fill}
\includegraphics[scale=0.28]{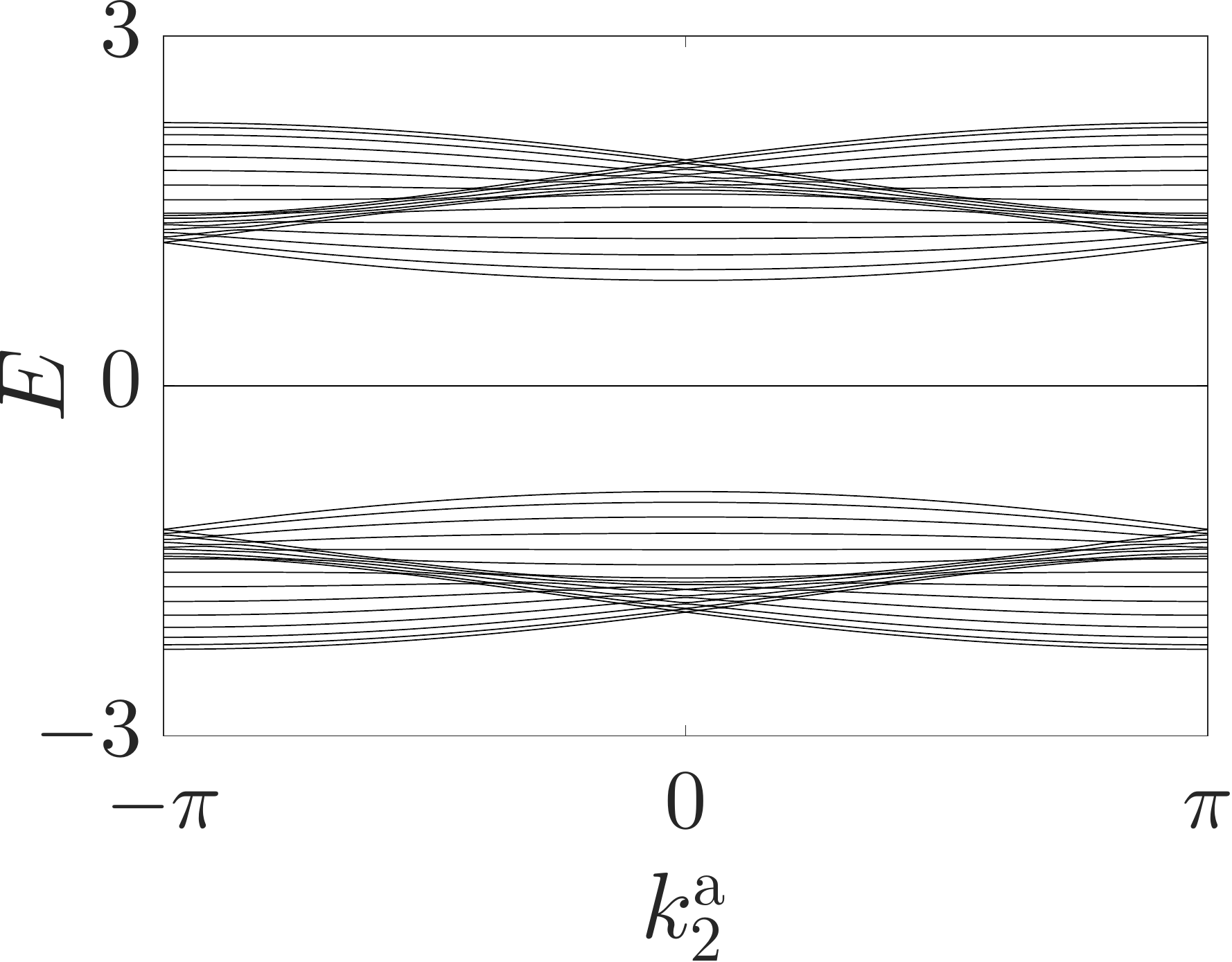}
\hspace*{\fill}
\includegraphics[scale=0.28]{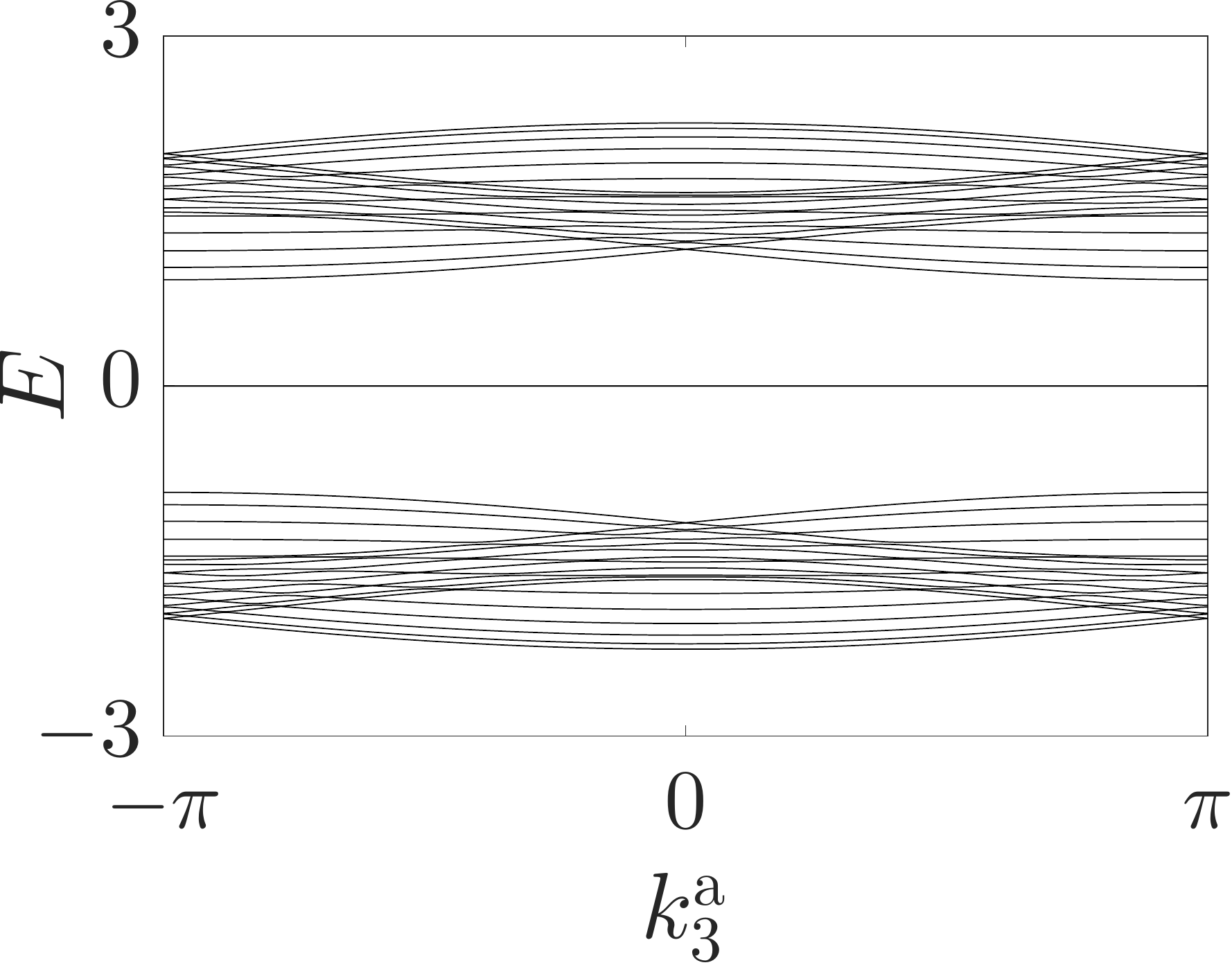}
\hspace*{\fill} \\[1ex]
\hspace*{\fill}
\includegraphics[scale=0.28]{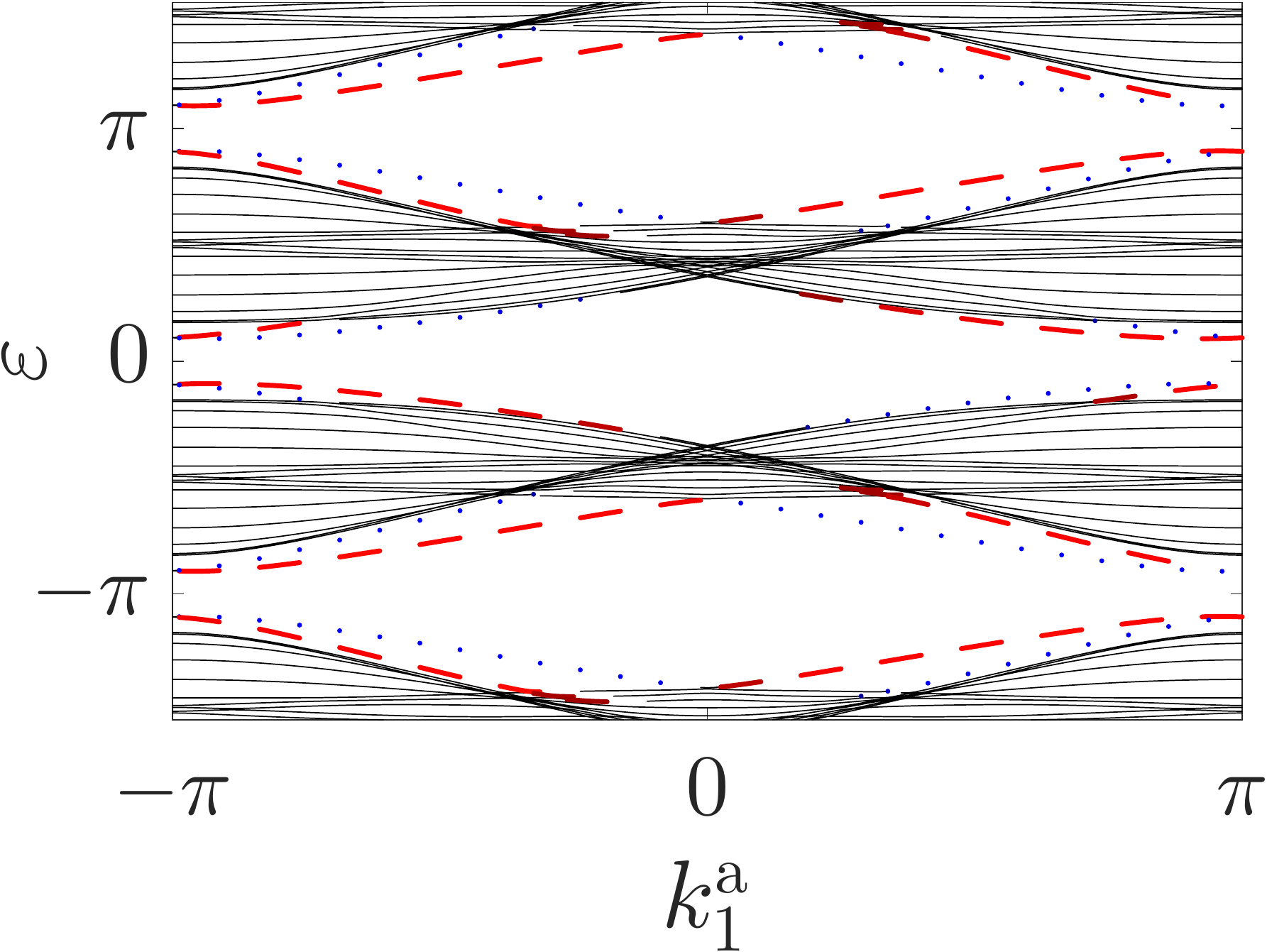}
\hspace*{\fill}
\includegraphics[scale=0.28]{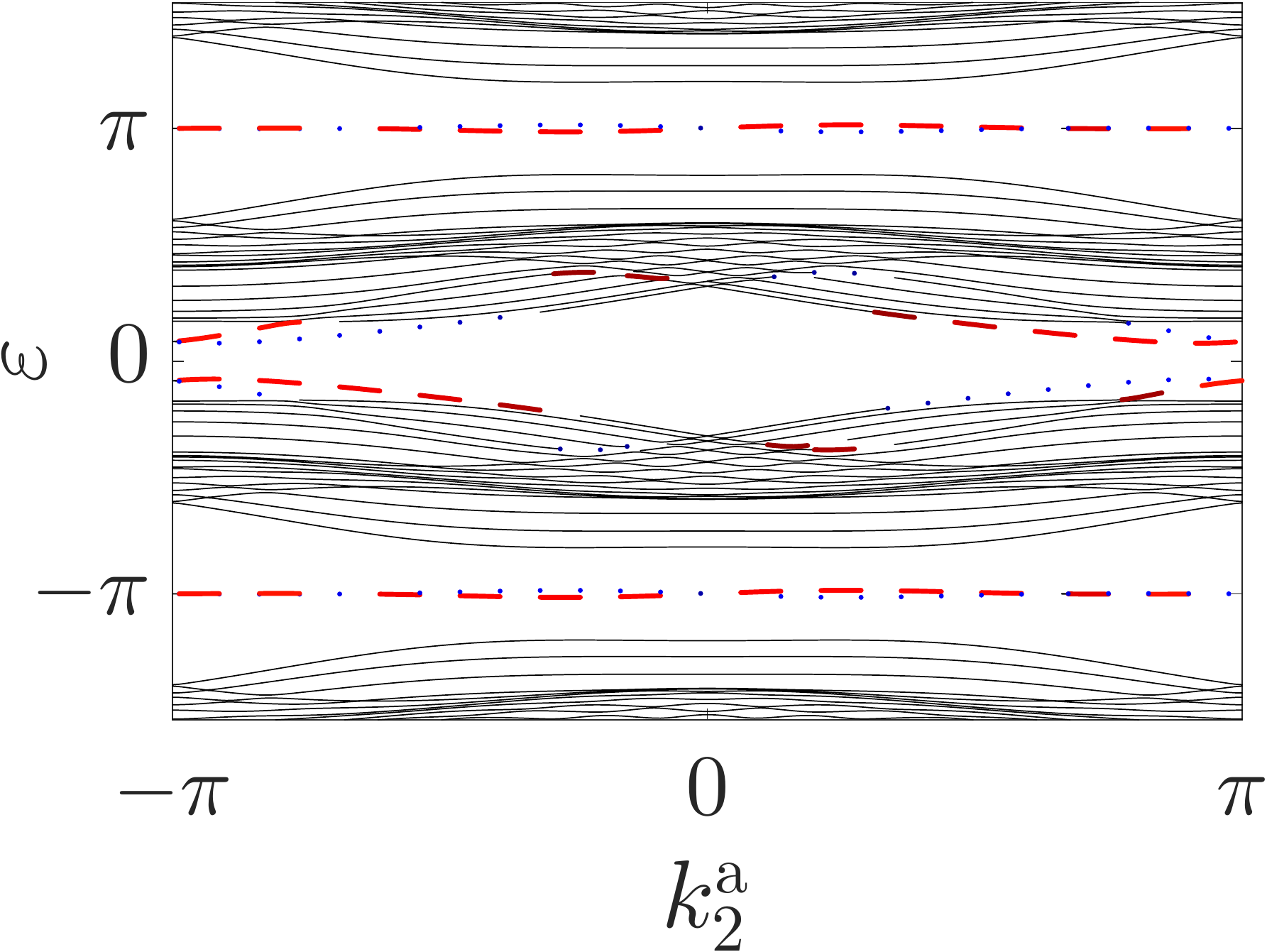}
\hspace*{\fill}
\includegraphics[scale=0.28]{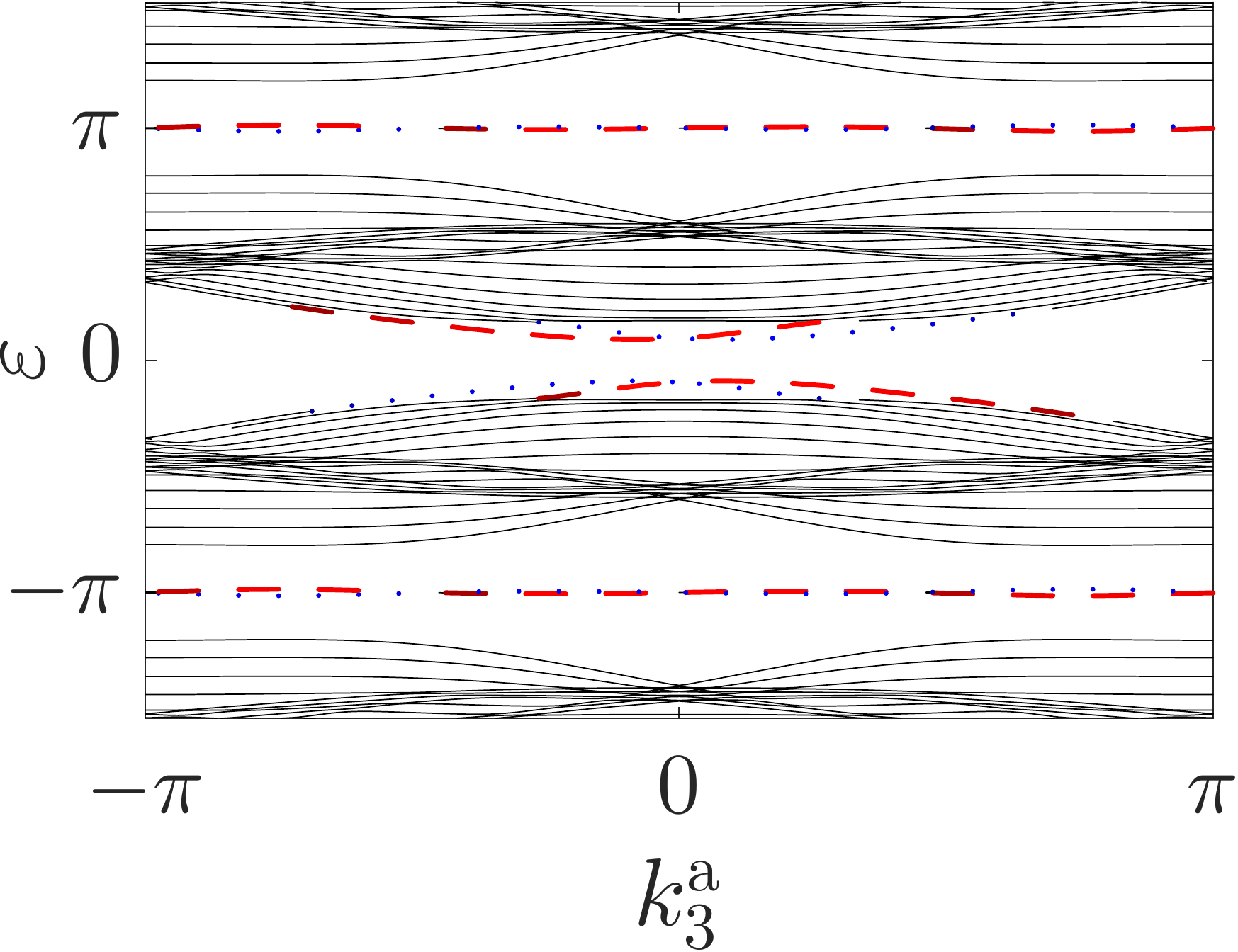}
\hspace*{\fill}
\caption{Same as Fig.~\ref{fig:zigzag}, now for armchair boundaries along the nearest-neighbor vectors $\boldsymbol \delta_1$, $\boldsymbol \delta_2$, $\boldsymbol \delta_3$, and momenta $k^a_1$, $k^a_2$, $k^a_3$.}
\label{fig:armchair}
\end{figure*}

To evaluate Eq.~\eqref{W_PH} we further need to determine the contribution $N^{\nu}(\epsilon, \vec d_i) \, C^{\nu}(\vec d_i)$ from unpaired degeneracy points at each IM $\vec M_0, \dots, \vec M_3$.
For the model from Eq.~\eqref{PH_Ham}, in the situation of Fig.~\ref{fig:PH} and Figs.~\ref{fig:zigzag},~\ref{fig:armchair}, 
these values are given in Tab.~\ref{tab:PH1}.
They have been determined from the propagator $U(\vec k, t)$ for $0 \le t \le T$, using the algorithm from Ref.~\onlinecite{HAF17}.

\begin{table}[b]
\begin{center}
\begin{ruledtabular}
\begin{tabular}{ccccc}
$\epsilon$ & $\vec M_0$ & $\vec M_1$ & $\vec M_2$ & $\vec M_3$ \\\hline
 $0$  & $0$ & $0$ & $1$ & $1$  \\
 $\pi$ & $0$ & $0$ & $1$ & $1$
\end{tabular}
\end{ruledtabular}
\end{center}
\caption{Values of $N^{\nu}(\epsilon, \vec d_i) \, C^{\nu}(\vec d_i)$ for Figs.~\ref{fig:zigzag},~\ref{fig:armchair}.}
\label{tab:PH1}
\end{table}

With the information from Tab.~\ref{tab:PH1} we can now immediately evaluate Eq.~\eqref{W_PH}.
In the gap at $\epsilon=\pi$ we obtain the $W_\mathrm{ph}$-invariants given in the first column of Tab.~\ref{tab:PH2}.
Note that because of $W_3(\pi) = 0$ we have $W_\mathrm{ph}^0(\pi) = W_\mathrm{ph}^\pi(\pi)$.
Comparison with Figs.~\ref{fig:zigzag},~\ref{fig:armchair}, where the boundary states are shown explicitly, confirms the correctness of the $W_\mathrm{ph}$-invariants.

\begin{table}
\begin{center}
\begin{ruledtabular}
\begin{tabular}{l|c|ccc}
 & $W_\mathrm{ph}^{0,\pi}(\pi)$ & $N(t=0)$ & $\sum N^\nu C^\nu$ & $W_\mathrm{ph}^{0,\pi}(0)$ \\\hline
 $\vec a_1$ & $0$ & $1$ & $0$ & $1$  \\
 $\vec a_2$ & $1$ & $0$ & $1$ & $1$  \\
 $\vec a_3$ & $1$ & $0$ & $1$ & $1$  \\[0.5ex]
 $3\boldsymbol \delta_1$ & $0$ & $0$ & $0$ & $0$  \\
 $3\boldsymbol \delta_2$ & $1$ & $1$ & $1$ & $0$ \\
 $3\boldsymbol \delta_3$ & $1$ & $1$ & $1$ & $0$ \\
\end{tabular}
\end{ruledtabular}
\end{center}
\caption{$W_\mathrm{ph}$-invariants for Figs.~\ref{fig:zigzag},~\ref{fig:armchair}.}
\label{tab:PH2}
\end{table}

In the gap at $\epsilon=0$ another complication arises due to the possibility of boundary states for $t=0$.
In the upper rows of Figs.~\ref{fig:zigzag},~\ref{fig:armchair} we show the boundary spectrum of the initial Hamiltonian $H_\mathrm{ph}(t=0)$, which is the starting point for the subsequent evolution described by $U(\cdot)$.
Depending on the boundary orientation, $H_\mathrm{ph}(t=0)$ can possess a boundary state at $\epsilon=0$.
In the present situation, where $H_\mathrm{ph}(t=0)$ is particle-hole and (as a real-valued Hamiltonian) time-reversal symmetric, the dispersion of the boundary state is perfectly flat. Recall that $U(\vec k, t)$, on the other hand, is not time-reversal symmetric according to Eq.~\eqref{TR_symm}.

The initial boundary states must be included in Eq.~\eqref{W_PH},
just as we had to do for the $W_3$-invariant in Eq.~\eqref{W3_new} if the bands are not topologically trivial at $t=0$.
Any initial boundary state changes the corresponding $W_\mathrm{ph}$-invariant by one, that is, through counting modulo two, flips its value between zero and one.

The number of initial boundary states $N(t=0)$ in Tab.~\ref{tab:PH2} can be taken from the upper rows in Figs.~\ref{fig:zigzag},~\ref{fig:armchair}. 
The contribution from the degeneracy points of $U(\vec k, t)$ is given in the third column of this table.
Summation of both numbers now gives the $W_\mathrm{ph}$-invariants for the gap at $\epsilon = 0$.
Note that because of $W_3(0) = 0$ we have again $W_\mathrm{ph}^0(0) = W_\mathrm{ph}^\pi(0)$.
Comparison with Figs.~\ref{fig:zigzag},~\ref{fig:armchair} confirms the correctness of the $W_\mathrm{ph}$-invariants, also in cases where initial boundary states have to be taken into account.

In the main text, for Fig.~\ref{fig:PH}, we have selected two boundaries without initial boundary states (namely, the third column from Fig.~\ref{fig:zigzag} and the first column in Fig.~\ref{fig:armchair}),
which allowed for a straightforward discussion.
With the present results for all boundaries, we recognize the full complexity associated with the `weak' topological phase.

For the gap at $\epsilon=\pi$, initial boundary states do not play a role (they simply do not exist outside of the spectrum of $H_\mathrm{ph}(t=0)$). According to the $0$-$\pi$ pattern of the projections $\vec a_i \cdot \vec M_m$ or  $\boldsymbol \delta_i \cdot \vec M_m$, we expect that in a `weak' phase symmetry-protected boundary states exist for two out of three boundary orientations.
This is true for both zigzag and armchair boundaries in Figs.~\ref{fig:zigzag},~\ref{fig:armchair}.

For the gap at $\epsilon=0$, the ``two-out-of-three'' rule does not apply because of the initial boundary states.
For armchair boundaries, symmetry-protected boundary states do not occur for any boundary orientation.
For zigzag boundaries, symmetry-protected boundary states occur for every boundary orientation.
We like to stress that this effect is not a simple consequence of the different geometry of armchair and zigzag boundaries.
In particular, as Figs.~\ref{fig:zigzag},~\ref{fig:armchair} show, no immediate relation between the appearance of boundary states at $t=0$ and at $t=T$ exists.
 Unless one computes the full $W_\mathrm{ph}$-invariants,
 which keep track of the creation and annihilation of symmetry-protected states during time-evolution,
 the entire situation remains obscure.
  
\vspace*{-1ex}

\end{document}